\documentclass[11pt,a4paper,times]{article}

\usepackage{import}
\usepackage[margin=1.1in]{geometry} 
\usepackage[utf8]{inputenc}
\usepackage[english]{babel}
\usepackage{csquotes}
\usepackage{graphicx}
\usepackage{amsmath}
\usepackage{booktabs}
\usepackage{lscape}
\usepackage{longtable}
\usepackage{caption}
\usepackage{authblk}
\usepackage{bm}
\usepackage[unicode=true]{hyperref} 
\hypersetup{colorlinks=true,linkcolor=blue,urlcolor=blue,citecolor=blue}
\usepackage{caption} 
\usepackage[backend=bibtex,style=numeric,giveninits=true,maxbibnames=99,sorting=none]{biblatex} 
\title{Trends in COVID-19 hospital outcomes in England before and after vaccine introduction, a cohort study}
\author[1]{PD Kirwan}
\author[2]{A Charlett}
\author[2]{P Birrell}
\author[2]{S Elgohari}
\author[2]{R Hope}
\author[2]{S Mandal}
\author[1,2,*]{D De Angelis}
\author[1,*]{AM Presanis}
\affil[1]{Medical Research Council Biostatistics Unit, School of Clinical Medicine, University of Cambridge, Cambridge, UK}
\affil[2]{UK Health Security Agency, London, UK}
\affil[*]{These authors jointly supervised this work.}
\date{\vspace{-5ex}}

\begin{document}

\maketitle

\section*{Abstract}

Widespread vaccination campaigns have changed the landscape for COVID-19, vastly altering symptoms and reducing morbidity and mortality. We estimate trends in mortality by month of admission and vaccination status among those hospitalised with COVID-19 in England between March 2020 to September 2021, controlling for demographic factors and hospital load.

Among 259,727 hospitalised COVID-19 cases, 51,948 (20.0\%) experienced mortality in hospital. Hospitalised fatality risk ranged from 40.3\% (95\% confidence interval 39.4-41.3\%) in March 2020 to 8.1\% (7.2-9.0\%) in June 2021. Older individuals and those with multiple co-morbidities were more likely to die or else experienced longer stays prior to discharge. Compared to unvaccinated people, the hazard of hospitalised mortality was 0.71 (0.67-0.77) with a first vaccine dose, and 0.56 (0.52-0.61) with a second vaccine dose. Compared to hospital load at 0-20\% of the busiest week, the hazard of hospitalised mortality during periods of peak load (90-100\%), was 1.23 (1.12-1.34).

The prognosis for people hospitalised with COVID-19 in England has varied substantially throughout the pandemic and according to case-mix, vaccination, and hospital load. Our estimates provide an indication for demands on hospital resources, and the relationship between hospital burden and outcomes.

\newpage

\section*{Introduction}
It is now well established that a segment of the population who acquire severe acute respiratory syndrome coronavirus 2 (SARS-CoV-2) infection, the virus responsible for coronavirus disease 2019 (COVID-19), in the community will require hospitalisation, potential escalation to intensive care facilities, and may die in hospital or soon after discharge. COVID-19 has been shown to disproportionately impact older people and those with multiple comorbidities, compared to younger, healthier individuals, and there is considerable evidence that these factors heavily influence prognosis following hospital admission for COVID-19 [1–4]. 

The extensive vaccination campaign in England during 2021 has dramatically changed the outlook for COVID-19, lessening symptoms and reducing morbidity and mortality [5,6]. Despite widespread and high levels of vaccination, however, individuals continue to experience COVID-19 infection severe enough to require hospitalisation. Several studies have previously examined COVID-19 hospitalised fatality risk (HFR) in England according to baseline demographic factors [1–3,7–9] but there is little information about prognosis in the current context of vaccination across the population and how this might be related to hospital load.

We aimed to investigate trends in mortality within 90 days of hospitalisation with COVID-19 among a national cohort of all people hospitalised with community-onset COVID-19 in England and how these trends vary according to vaccination status, hospital load, and other factors. We apply statistical methods which account for competing outcomes to estimate absolute and relative risks of hospitalised fatality, and lengths of stay in hospital by outcome, and control for or assess the potential impact of different biases.

\newpage

\section*{Results}

\subsection*{Participant characteristics}

Among 259,727 people with COVID-19 hospitalised between 15th March 2020 and 30th September 2021, a total of 51,948 (20.0\%) died, 191,663 (73.8\%) were discharged and the remaining 16,116 (6.2\%) remained in hospital at the date of data extraction and/or were right-censored at 90 days (Table 1). Figure 1 presents weekly hospital admissions for COVID-19 over the study period, with an indication of the first, second, and third waves. Figure 2 shows the vaccination status of the study population between December 2020 and September 2021, by month of admission and age group, demonstrating the strong correlation between age and vaccination during the second and third waves.

Table 1 presents patient characteristics for the study population. Compared to all people with PCR-confirmed community-acquired COVID-19 infection, those hospitalised for COVID-19 were older (48.4\% aged over 65 vs. 10.0\%), more likely to be male (52.1\% vs. 47.6\%), and to reside in London (18.0\% vs. 16.3\%). A greater proportion of those hospitalised were of Black ethnicity (5.9\% vs 4.2\%) and lived in an area of high deprivation (28.1\% vs 23.6\%), compared to all those with community-acquired COVID-19. Comparative information on comorbidity was not available, although 7.2\% of those hospitalised had a Charlson comorbidity index (CCI) score of 5 or more (Table 1) compared to 2.3\% among a sample of 657,264 individuals 20 years and older registered at English primary care practices in 2005 [10].

\subsection*{Observed hospital outcomes}

In unadjusted comparisons, older individuals experienced poorer outcomes following hospital admission; almost half (46.5\%) of those aged 85+ died in hospital compared to just 0.5\% of those aged 15-24. Similarly, males (as compared to females), those living outside of London and the South West, and those with an increased CCI score (compared to lower CCI) were more likely to die in hospital (Supp. Table 1).

\subsection*{Hospitalised case-fatality risk}

Figure 3, panel a shows estimated HFR by month of hospital admission. HFR decreased during the first wave of the pandemic from 40.3\% (95\% confidence interval: 39.4-41.3\%) in March 2020 to 12.3\% (10.3-14.8\%) in August 2020. During the second wave HFR increased to a peak of 22.8\% (22.5-23.2\%) in January 2021, although by March 2021 had halved to 10.5\% (9.7-11.4\%) and has remained at or below 10\% throughout subsequent months (Supp. Table 2).

In estimates of HFR by month and a single other covariate i.e. unadjusted for all other covariates older individuals had increased HFR. In March 2020 HFR was 1.4\% (0.4-5.7\%) among those aged 0-14, 22.3\% (20.9-23.9\%) for those aged 45-64 and 66.5\% (64.4-68.7\%) among those aged 85+ (Figure 4, panel a, Supp. Table 3). Males tended to have greater HFR than females, although the extent of this differed by month, being more evident in March 2020 (HFR of 43.7\% (42.5-44.9\%) among males compared to 35.2\% (33.8-36.6\%) among females) than in April 2021 (HFR of 8.6\% (6.9-10.6\%) among males compared to 8.9\% (7.2-10.9\%) among females) (Supp. Figure 1, Supp. Table 4). HFR was lower for those residing in London and the South West: in December 2020 HFR was 18.7\% (17.8-19.6\%) in London and 17.5\% (15.8-19.4\%) in the South West, with a point estimate above 22\% in all other regions of England (Supp. Figure 3, Supp. Table 6). Lastly, HFR was increased among those with a higher CCI; in December 2020 HFR was 6.7\% (6.2-7.2\%) among those with a CCI of 0, 22.4\% (21.6-23.1\%) with a CCI of 1-2, 38.7\% (37.3-40.2\%) with a CCI of 3-4, and 47.3\% (45.3-49.4\%) with a CCI of 5 or above. This association was seen throughout the study period (Supp. Figure 4, Supp, Table 7).

During the initial months of the study those admitted to hospitals which were experiencing higher load had a greater HFR as compared to those admitted to hospitals with lower activity e.g. during March 2020, HFR was 44.9\% (39.6-50.9\%) for hospitals at 90-100\% of their peak load, compared to 38.8\% (36.6-41.0\%) for hospitals at 0-20\% of their peak load. This disparity appeared to lessen during the summer months, but was seen again during autumn of 2020, as admissions began rising. In November 2020, HFR was 23.6\% (21.1-26.4\%) for hospitals at 90-100\% of peak load, compared to 17.4\% (16.4-18.6\%) for hospitals at 0-20\% (Supp. Table 8).

\subsection*{Length of stay}

Figure 3, panel b shows estimated median lengths of stay until death or discharge by month of hospital admission. Aside from the first two months of the pandemic (March-April 2020), those with an eventual outcome of death had longer stays in hospital compared to those who were discharged. The length of stay prior to death and discharge followed approximately inverse trends: whilst length of stay prior to discharge decreased throughout the first wave, from 5.9 (5.8-6.0) days in March 2020 to 3.1 (2.9-3.3) days in August 2020, length of stay prior to death increased from 5.6 (5.5-5.6) days to 9.9 (9.2-10.9) days.

During the second wave, lengths of stay prior to discharge initially lengthened, peaking at 5.0 (5.0-5.1) days in December 2020, before falling to 2.7 (2.6-2.8) days by June 2021. Conversely, length of stay prior to death was shortest in January, at 7.5 (7.5-7.6) days, and lengthened to 10.4 (10.1-10.8) days by June 2021 (Supp. Table 2). 

Examining the estimated lengths of stay for different subgroups (Figure 4, panel b, Supp. Figures 1-5, Supp. Tables 3-8), similar patterns were observed for males and females, although with less pronounced variation among males. Length of stay prior to death estimates were imprecise for younger individuals, due to the small number of events, but for age groups 45-64 and above, the median length of stay prior to death decreased with increasing age. Meanwhile older individuals had longer stays in hospital prior to discharge compared to younger individuals; the median time to discharge among those aged 85+ ranged between 5.1-10.4 days, compared to between 0.8-2.4 days for those aged 0-14 and 15-24 (Figure 4, panel b). Those with a higher CCI similarly experienced shorter stays prior to death and longer stays prior to discharge; in December 2020 those with a CCI above 5 remained in hospital for a median of 7.0 (6.8-7.2) days prior to death and 8.9 (8.7-9.0) days prior to discharge, compared to 10.6 (10.3-11.0) days and 3.4 (3.3-3.4) days for those with a CCI of 0. Hospitals experiencing higher load had similar lengths of stay until discharge but shorter lengths of stay until death as compared to those with lower activity.

\subsection*{HFR by vaccination status}

Table 2 shows HFR by vaccination status and age group. For all ages, vaccination was associated with a reduced HFR, with significant reductions in HFR among those hospitalised $\geq$21 days post first vaccine or $\geq$14 days post second vaccine. The HFR for a double vaccinated adult aged 75-85 was 22.5\% (20.4-24.8\%), this compares to 38.6\% (37.7-39.6\%) for an unvaccinated adult in the same age group and 25.3\% (24.5-26.0\%) for an unvaccinated adult aged 65-75.

\subsection*{Relative risks}

Figure 5 presents hazard ratios for hospitalised fatality by month of admission. Controlling for age group, region of residence, vaccination status, sex, ethnicity, index of multiple deprivation (IMD) quintile, CCI and hospital load, month of hospital admission remained a significant factor for the prognosis of hospitalised individuals. Compared to June 2020, the hazard for hospitalised fatality was increased during March to May 2020, September 2020 to February 2021, and June to September 2021 (Figure 5, Supp. Table 9).

Similarly, controlling for month of admission and the factors mentioned above, vaccination status was a significant factor for prognosis following hospitalisation. During January to September 2021, compared to the reference category of unvaccinated, the hazard ratio for hospitalised fatality was 0.93 (0.89-0.98) for individuals hospitalised <21 days after first vaccination dose, 0.71 (0.67-0.77) for individuals hospitalised $\geq$21 days after first vaccination dose, and 0.56 (0.52-0.61) for individuals hospitalised $\geq$14 days after second vaccination dose (Figure 6, Supp. Table 10).

There was an increased hazard of hospitalised fatality for those of Asian (1.19 (1.13-1.25)) but reduced for those of Black ethnicity (0.90 (0.84-0.97)) compared to reference category White. Males had a greater hazard compared to females (1.28 (1.24-1.32)), and compared to a CCI of 0, those with a higher burden of comorbidity had a greater hazard of hospitalised fatality (3.46 (3.27-3.67) for CCI of 5 and above). Those residing in more deprived quintiles had greater hazards for hospitalised fatality (1.10 (1.05-1.15) for the most deprived quintile) compared to a reference of least deprived (Supp. Figure 6, Supp. Table 10).

Hazards were also elevated with increased hospital load, up to 1.23 (1.12-1.34) for load at 90-100\% of the busiest week (compared to 0-20\% load).

\subsection*{Sensitivity to epidemic phase bias}

Supplementary Figure 7 shows the outcome of the shift sensitivity analyses by month of symptom onset, adjusted for the same covariates as above. The greatest effect was observed for the March 2020 hazard ratio estimate, which steadily reduced towards 1 following a shift of c=1, 2, 3 or 4 days. The effect in other months was small, with the previously described monthly trends persisting, although the slight reduction in hazard estimated for the most recent month (September 2021) was no longer apparent.

\newpage

\section*{Discussion}
People continue to experience hospitalisation for severe COVID-19, we aimed to investigate how hospital prognosis and length of stay has changed with the advent of vaccination and in the context of varying hospital pressures. We examined absolute and relative risks of hospitalised fatality and lengths of stay in hospital during the first year and a half of the COVID-19 pandemic in England. 

\subsection*{Findings in context}

We identified a number of studies exploring COVID-19 HFR in England according to demographic factors [1–3,7,11–13]. In line with these epidemiological studies, we found that people with community-acquired COVID-19 who became hospitalised were older, more likely to be male, of Black ethnicity, and to live in areas of high deprivation, as compared to everyone diagnosed with the virus. Among those who were hospitalised, we estimated greater absolute fatality risks for men and older individuals, and HFR also varied according to ethnicity, month of admission, hospital load, and region. Lengths of stay in hospital were similarly associated with demographic factors, with median lengths of stay prior to death typically longer than those prior to discharge. In relative risk analyses controlling for all measured confounders, baseline comorbidity burden was the strongest predictor of death.

Prior to this study there was limited information available on COVID-19 outcomes at English hospitals by hospital load, although a recent King’s Fund study concluded that a shortage of overnight and acute bed availability prior to the pandemic had already put hospitals under increased strain [14]. In Switzerland meanwhile, increased hospital load has been associated with poorer outcomes for COVID-19, with an ICU occupancy of 70\% or greater estimated to be a tipping point at which outcomes became adversely affected [15]. 

Our estimates suggest a deterioration in survival as hospital load increases, however, there are several potential biases which make this finding hard to interpret. During periods of peak hospital load there is likely a modification of an individual’s willingness to attend healthcare services for mild illness, changes in admission criteria both to wards and intensive care units [16], and individuals with milder disease may be selected for transfer from overloaded hospitals to those with bed availability, due to the lower inherent risk. Each of these factors may influence the case-mix at times of peak hospital demand.

There is now compelling evidence that vaccination reduces the number of individuals being hospitalised [5] and the risk of mortality, regardless of hospital admission [9,17]. We found reduced hospitalised fatality among vaccinated individuals, with the reduction most clearly seen among older individuals. For those aged 75 and over, vaccination reduced HFR to approximately the risk of an unvaccinated individual aged 10 years younger. In adjusted estimates, each additional vaccine dose reduced the hazard for fatality by a significant margin, with a 42\% (38-46\%) estimated reduction in the risk of death for double-vaccinated individuals. This is a slightly lower reduction than for all community-acquired PCR-positive COVID-19 cases in England, where a 51\% (37-62\%) reduced risk of death was estimated for symptomatic individuals who had received a single vaccine [5]. This difference may reflect the portion of hospitalised individuals who die from other causes, or could be an indication of waning vaccine efficacy among our study population.

After controlling for all measured covariates, including hospital load and vaccination, we continued to estimate monthly variation in outcomes, with apparent seasonal variation in hazards. Whilst seasonal patterns in respiratory pathogens such as influenza and respiratory syncytial virus are well-documented [18,19], a multitude of interlinked factors including changes in national restrictions and the emergence of new variants may have influenced these trends.

\subsection*{Strengths and limitations}

The use of high-quality hospital surveillance data linked to several other comprehensive data sources is a strength of this study and enabled a broader understanding of the factors influencing hospitalised fatality. For covariates with varying levels of completeness we undertook sensitivity analyses to confirm minimal effects on our estimates (e.g. indication of injury as a factor for emergency care admission), however, there may have been other unmeasured confounders for which we could not account. Using appropriate statistical methods we adjusted for competing risks, and the use of a relatively coarse monthly timescale likely limited the extent to which our study was affected by epidemic phase bias [20]. Data linkage allowed for deaths occurring shortly after discharge to be identified, almost a fifth of all deaths in the study occurred within 14 days of discharge, suggesting palliative discharge.

Data on hospital pathways following admission were unavailable. As such, we were not able to subdivide the hospitalised population by severity of infection and/or need of respiratory support, whether within or outside of intensive care. Treatment data and changes in patient management were similarly unmeasured in our dataset, although the use of therapeutic agents is likely to have contributed to the reduction in hospital fatality risk, particularly at the start of the pandemic [7,21].

The measure of hospital burden we used considered acute hospital admissions at and around the time of admission as a proxy for bed occupancy. Whilst no single accepted measure of hospital burden exists, overnight bed occupancy is a widely used metric [14], and guidance on bed occupancy was issued to ICUs (e.g. alterations in practice upon reaching 150\% and 200\% above pre-pandemic baseline) [16]. A limitation of the bed occupancy measure is that it only measures demand and not supply (i.e. staffing levels), or the extent of other hospital pressures. Work to access and integrate measures of supply is ongoing.

Lastly, this study did not consider the significant proportion of individuals (up to 40\%) who may have acquired COVID-19 nosocomially (in hospital). Fatality risks and lengths of hospital stay for these individuals are complicated by other conditions. Whilst not considered in our estimates, researchers in Scotland have found similar effects of age, sex, and comorbidity upon prognosis following nosocomial COVID-19 acquisition [22].

\subsection*{Summary}

Hospital outcomes and lengths of stay continue to vary according to case-mix, vaccination, and changes in hospital load more than 18 months after the pandemic began in England. One of the primary goals of the lockdown measures implemented in England at various times since the start of the pandemic has been to protect against hospitals becoming excessively overburdened. Even with these measures in place, being admitted during a period of high hospital load was correlated with poorer outcomes. Meanwhile, vaccinated individuals admitted to hospital for COVID-19 had a significantly reduced risk of mortality.

Outcomes following hospitalisation with COVID-19 should continue to be monitored, particularly with the emergence of new variants. The datasets and methods we describe continue to be vital to estimate changes in severity, providing an indication for demands on hospital resources, resulting effects on waiting lists for elective procedures, and monitoring the relationship between hospital burden and outcomes.

\newpage

\section*{Methods}

\subsection*{Study design and setting}

A retrospective cohort study using competing risk regression to estimate relative risks of severe outcomes. We consider data on hospital admissions in England since the initial wave of COVID-19 in March 2020 until the end of September 2021, with follow-up until 22nd November 2021. 

\subsection*{Participants}

All individuals aged 15 years and older with community-acquired COVID-19 (defined as a positive test for COVID-19 within -14 to 1 days of hospital admission), admitted to hospital in England for COVID-19 or another non-injury related condition between 1st March 2020 and 30th September 2021 were included (n=259,727). Hospital records with inconsistent date information (n=2) or missing demographic information (n=302) were excluded.

\subsection*{Data sources and outcomes}

The United Kingdom Health Security Agency (UKHSA), alongside NHS England, monitors infectious diseases in England. The NHS England Secondary Uses Service (SUS) dataset contains well completed, accurate information on hospitalisation for COVID-19 in England, along with identifiers to augment these data through linkage to other routinely collected information. Admissions are, however, only entered into the SUS dataset upon completion of a hospital stay (i.e. at the point of discharge from hospital or death). So the information in SUS data was supplemented with information on individuals still in hospital though linkage to the Emergency Care Dataset for England (ECDS), which promptly records all emergency care attendances and onwards destinations (i.e. discharge home or admittance to hospital). Among those with completed hospital episodes, 77\% were admitted via emergency care.

Complete information on deaths was obtained through linkage to the UKHSA deaths dataset, containing all dates of death for people with a positive COVID-19 test. Date of vaccination (first and second dose, third doses were not considered as only n=11 hospitalised case had received a third dose during the follow-up period) was obtained through linkage to the UKHSA National Immunisation Management Service (NIMS). Testing information both for community-acquired COVID-19 infections identified through PCR testing on arrival at hospital (Pillar 1) and PCR testing within the community (Pillar 2) were obtained from the UKHSA Second Generation Surveillance System (SGSS). All data were stored and analysed on UKHSA computers under agreed data governance protocols.

\subsection*{Covariates}

Covariates in the linked dataset included vaccination status (unvaccinated, <21 days of first dose, $\geq$21 days after first dose, $\geq$14 days after second dose), date of hospital admission (aggregated by month), age group, region of residence (Government office region), Charlson Comorbidity Index (CCI) [23], ethnicity, sex, index of multiple deprivation (IMD) quintile, and measure of hospital load. The hospital load measure was defined as the number of COVID-19 admissions at an NHS trust within the 7 days around admission (3 before, same day, and 3 after), as a proportion of the busiest 7-day period at that trust. Hospital load was grouped into: 0-20\%, 20-40\%, 40-60\%, 60-80\%, 80-90\%, and 90-100\%. In relative risk analyses the two key exposure variables considered were vaccination status and month of hospital admission.

\subsection*{Representativeness}

Data comprised all new admissions for COVID-19 reported in England. Numbers of reported admissions were compared with the NHS weekly COVID-19 admissions data [24] to ensure data were representative. Hospital-onset COVID-19 (i.e. infection occurring in hospital) cases were excluded: those who acquired a hospital-onset infection during the study period (n=208,851) tended to be older and have longer lengths of stay than the community-onset cases considered in this study, with a greater proportion remaining in hospital post-90 days (Table 1). Hospital stays for these individuals may be influenced by conditions other than COVID-19, as described by Bhattacharya et al [25].

\subsection*{Bias}

Data validation was undertaken between the linked datasets, we found no systematic under-reporting or misreporting of person characteristics and linked information was used to minimise missing data. Censored outcomes and competing risks were explicitly accounted for by the choice of statistical method.

We carried out a sensitivity analysis to assess the potential effect of epidemic phase bias on the estimated hazard ratios in relative risk analyses [20]. This bias is caused by conditioning on an observed date later than the date of infection (see Supplementary Information for a full description), therefore in the sensitivity analysis we conditioned on date of symptom onset, which is nearer to date of infection than date of hospital admission. This conditioning ensured that the sensitivity analysis targeted bias due to epidemic phase, as opposed to any other factors which may influence time from symptom onset to admission.

\subsection*{Statistical methods}

In our study, hospitalised individuals are at risk of more than one event during the follow-up period i.e. they can die or be discharged. In this competing risk context standard survival analysis may result in biased estimates of the absolute and relative risks of hospitalised fatality, particularly when one of the competing risk is large (e.g. discharges to palliative care as a competing risk for death) [26]. Therefore, two alternative, complementary, statistical analyses were undertaken (see Supplementary Information for more details). 

We used Aalen-Johansen cumulative incidence estimation to obtain estimates of cumulative HFR and median lengths of stay in hospital for specific sub-sets of the population, unadjusted for other factors [27]. Median lengths of stay were the weighted median estimate with weighted ties. We used stratified Fine-Gray competing risk regression with adjustment for confounders to estimate the association of each risk factor with the cumulative incidence of mortality within 90 days of hospitalisation with COVID-19 (we term this hospitalised fatality). Fine-Gray regression models the proportional sub-distribution hazard of hospitalised fatality derived from the cumulative incidence function [28]. Stratification was used for confounders with non-proportional hazards (see Supplementary Information).

\subsection*{Censoring}

To focus our analyses on outcomes following COVID-19 admission, a pragmatic cut-off of 90 days from first positive specimen date was chosen and only those hospital outcomes (death or discharge) occurring within this cut-off were included. All records with outcomes occurring beyond 90 days (n=656) were right-censored at 90 days, while individuals who remained in hospital at the date of data extraction (n=15,460) were right-censored at the shorter of this date or 90 days. To better account for palliative discharge, deaths occurring within 14 days of discharge from hospital (n=9,933, 19.1\% of all deaths observed) were classified as deaths rather than discharges and the date of death used as the outcome date. Linkage to the UKHSA deaths data enabled these post-hospital discharge deaths to be identified.

\subsection*{Model implementation}

Statistical models were implemented using R version 4.1.1 (R Foundation for Statistical Computing, Vienna, Austria) and the open-source packages survival v. 3.2-12 [29], and matrixStats v. 0.60.0 [30].

\subsection*{Ethics approval}

This study does not contain patient identifiable data. Consent from individuals involved in this study was not required. The mandatory surveillance systems used in this study, NHS England Secondary Uses Service (SUS), Emergency Care Dataset for England (ECDS), UKHSA deaths dataset, UKHSA National Immunisation Management Service (NIMS), and UKHSA Second Generation Surveillance System (SGSS), are approved by the Department of Health and Social Care. Data were collected with permissions granted under Regulation 3 of The Health Service (Control of Patient Information) Regulations 2002, and without explicit patient permission under Section 251 of the NHS Act 2006.

\subsection*{Patient and public involvement}

This study was a retrospective cohort analysis. The research question, design and data collection were motivated by the response to an urgent public health emergency. The surveillance data were collected by NHS England and the UK Health Security Agency with permissions granted under Regulation 3 of The Health Service (Control of Patient Information) Regulations 2002, and without explicit patient permission under Section 251 of the NHS Act 2006. Although patients were not directly involved in the study design, the experiences of clinicians and public health officials interacting with patients informed the design of the data collection.

\subsection*{Dissemination to participants and related patient and public communities}

UKHSA and the MRC Biostatistics Unit have public facing websites and Twitter accounts @UKHSA and @MRC\_BSU. UKHSA and the MRC Biostatistics Unit engage with print and internet press, television, radio, news, and documentary programme makers.

\subsection*{Transparency statement}

The lead author affirms that this manuscript is an honest, accurate, and transparent account of the study being reported; that no important aspects of the study have been omitted; and that any discrepancies from the study as planned (and, if relevant, registered) have been explained.

\subsection*{Data availability}

The data used in this study are protected data. These data are not publicly available because the information is personal or special category personal data, and there is risk of ‘re-identification’ of data that has been anonymised by data matching, inference or deductive disclosure. 

Access to protected data is subject to robust governance protocols, where it is lawful, ethical and safe to do so. Individuals and organisations wishing to request access to data used in this study, from the NHS England Secondary Uses Service (SUS), Emergency Care Dataset for England (ECDS), UKHSA deaths dataset, UKHSA National Immunisation Management Service (NIMS), or UKHSA Second Generation Surveillance System (SGSS) can make a request directly to NHS Digital (\url{https://digital.nhs.uk/services/data-access-request-service-dars}) or to UKHSA (\url{https://www.gov.uk/government/publications/accessing-ukhsa-protected-data}). Access to protected data is always strictly controlled using legally binding data sharing contracts.

Requests for underlying data cannot be granted by the authors because the data were acquired under licence/data sharing agreement from NHS Digital and UKHSA, for which conditions of use (and further use) apply.

The estimated or observed values underlying each figure can be found in the Supplementary tables.

\subsection*{Code availability}

Code for the survival and matrixStats R packages is available from the Comprehensive R Archive Network (\url{https://cran.r-project.org/}). R code written to process the data, implement the statistical analysis, and produce the figures and tables is available online [31].

\newpage

\section*{References}
\begin{enumerate}
\item Ferrando-Vivas P, Doidge J, Thomas K, Gould DW, Mouncey P, Shankar-Hari M, et al. Prognostic Factors for 30-Day Mortality in Critically Ill Patients With Coronavirus Disease 2019: An Observational Cohort Study. Crit Care Med. 2021;49: 102–111.
\item Kirwan PD, Elgohari S, Jackson CH, Tom BDM, Mandal S, De Angelis D, et al. Trends in risks of severe events and lengths of stay for COVID-19 hospitalisations in England over the pre-vaccination era: results from the Public Health England SARI-Watch surveillance scheme. arXiv [stat.AP]. 2021. Available: \url{http://arxiv.org/abs/2103.04867}
\item Navaratnam AV, Gray WK, Day J, Wendon J, Briggs TWR. Patient factors and temporal trends associated with COVID-19 in-hospital mortality in England: an observational study using administrative data. Lancet Respir Med. 2021. doi:10.1016/S2213-2600(20)30579-8
\item Agrawal U, Azcoaga-Lorenzo A, Fagbamigbe AF, Vasileiou E, Henery P, Simpson CR, et al. Association between multimorbidity and mortality in a cohort of patients admitted to hospital with COVID-19 in Scotland. J R Soc Med. 2021; 1410768211051715.
\item Lopez Bernal J, Andrews N, Gower C, Robertson C, Stowe J, Tessier E, et al. Effectiveness of the Pfizer-BioNTech and Oxford-AstraZeneca vaccines on covid-19 related symptoms, hospital admissions, and mortality in older adults in England: test negative case-control study. BMJ. 2021;373: n1088.
\item Public Health England. Direct and Indirect Impact of the Vaccination Programme on COVID-19 Infections and Mortality. 2021. Available: \url{https://assets.publishing.service.gov.uk/government/uploads/system/uploads/attachment_data/file/997495/Impact_of_COVID-19_vaccine_on_infection_and_mortality.pdf}
\item Docherty AB, Mulholland RH, Lone NI, Cheyne CP, De Angelis D, Diaz-Ordaz K, et al. Changes in UK hospital mortality in the first wave of COVID-19: the ISARIC WHO Clinical Characterisation Protocol prospective multicentre observational cohort study. bioRxiv. medRxiv; 2020. doi:10.1101/2020.12.19.20248559
\item Intensive Care National Audit Research Centre (ICNARC). ICNARC report on COVID-19 in critical care: England, Wales and Northern Ireland. 2021.
\item Thygesen JH, Tomlinson C, Hollings S, Mizani M, Handy A, Akbari A, et al. Understanding COVID-19 trajectories from a nationwide linked electronic health record cohort of 56 million people: phenotypes, severity, waves \& vaccination. bioRxiv. 2021. doi:10.1101/2021.11.08.21265312
\item Crooks CJ, West J, Card TR. A comparison of the recording of comorbidity in primary and secondary care by using the Charlson Index to predict short-term and long-term survival in a routine linked data cohort. BMJ Open. 2015;5: e007974.
\item Mathur R, Rentsch CT, Morton CE, Hulme WJ, Schultze A, MacKenna B, et al. Ethnic differences in SARS-CoV-2 infection and COVID-19-related hospitalisation, intensive care unit admission, and death in 17 million adults in England: an observational cohort study using the OpenSAFELY platform. Lancet. 2021;397: 1711–1724.
\item Gray WK, Navaratnam AV, Day J, Wendon J, Briggs TWR. Changes in COVID-19 in-hospital mortality in hospitalised adults in England over the first seven months of the pandemic: An observational study using administrative data. Lancet Reg Health Eur. 2021;5: 100104.
\item Public Health England. COVID-19: review of disparities in risks and outcomes. In: Gov.uk [Internet]. 2 Jun 2020 [cited 28 Jan 2022]. Available: \url{https://www.gov.uk/government/publications/covid-19-review-of-disparities-in-risks-and-outcomes}
\item Ewbank L, Thompson J, McKenna H, Anandaciva S, Ward D. NHS hospital bed numbers past, present, future. In: The King’s Fund [Internet]. 5 Nov 2021 [cited 16 Dec 2021]. Available: \url{https://www.kingsfund.org.uk/publications/nhs-hospital-bed-numbers}
\item Anderegg N, Panczak R, Egger M, Low N, Riou J. Mortality among people hospitalised with covid-19 in Switzerland: a nationwide population-based analysis. 2021. doi:10.31219/osf.io/37gaz
\item Advice on acute sector workforce models during COVID-19. NHS England; 2020 Dec. Available: \url{https://www.england.nhs.uk/coronavirus/publication/advice-on-acute-sector-workforce-models-during-covid-19/}
\item UK Health Security Agency. COVID-19 vaccine weekly surveillance reports (weeks 39 to 4, 2021 to 2022). In: Gov.uk [Internet]. 30 Sep 2021 [cited 28 Jan 2022]. Available: \url{https://www.gov.uk/government/publications/covid-19-vaccine-weekly-surveillance-reports}
\item Hardelid P, Pebody R. Mortality caused by influenza and respiratory syncytial virus by age group in England and Wales 1999–2010. Influenza Other Respi Viruses. 2013. Available: \url{https://onlinelibrary.wiley.com/doi/abs/10.1111/j.1750-2659.2012.00345.x}
\item Cromer D, van Hoek AJ, Jit M, Edmunds WJ, Fleming D, Miller E. The burden of influenza in England by age and clinical risk group: a statistical analysis to inform vaccine policy. J Infect. 2014;68: 363–371.
\item Seaman SR, Nyberg T, Overton CE, Pascall D, Presanis AM, De Angelis D. Adjusting for time of infection or positive test when estimating the risk of a post-infection outcome in an epidemic. 2021. doi:10.1101/2021.08.13.21262014
\item Beigel JH, Tomashek KM, Dodd LE, Mehta AK, Zingman BS, Kalil AC, et al. Remdesivir for the Treatment of Covid-19 — Preliminary Report. N Engl J Med. 2020;383: 1813–1836.
\item Khan KS, Reed-Embleton H, Lewis J, Saldanha J, Mahmud S. Does nosocomial COVID-19 result in increased 30-day mortality? A multi-centre observational study to identify risk factors for worse outcomes in patients with COVID-19. J Hosp Infect. 2021;107: 91–94.
\item Charlson ME, Pompei P, Ales KL, MacKenzie CR. A new method of classifying prognostic comorbidity in longitudinal studies: development and validation. J Chronic Dis. 1987;40: 373–383.
\item NHS England. COVID-19 Hospital Activity. [cited 21 Jun 2022]. Available: \url{https://www.england.nhs.uk/statistics/statistical-work-areas/covid-19-hospital-activity/}
\item Bhattacharya A, Collin SM, Stimson J, Thelwall S, Nsonwu O, Gerver S, et al. Healthcare-associated COVID-19 in England: A national data linkage study. J Infect. 2021;83: 565–572.
\item Andersen PK, Geskus RB, de Witte T, Putter H. Competing risks in epidemiology: possibilities and pitfalls. Int J Epidemiol. 2012;41: 861–870.
\item Aalen OO, Johansen S. An Empirical Transition Matrix for Non-Homogeneous Markov Chains Based on Censored Observations. Scand Stat Theory Appl. 1978;5: 141–150.
\item Fine JP, Gray RJ. A proportional hazards model for the subdistribution of a competing risk. J Am Stat Assoc. 1999. Available: \url{https://www.tandfonline.com/doi/abs/10.1080/01621459.1999.10474144}
\item Therneau T. A package for survival analysis in R. R package. Available: \url{https://github.com/therneau/survival}
\item Bengtsson H. matrixStats: Functions that Apply to Rows and Columns of Matrices (and to Vectors). R package. Available: \url{https://github.com/HenrikBengtsson/matrixStats}
\item Kirwan PD, Charlett A, Birrell P, Elgohari S, Hope R, Mandal S, et al. Trends in COVID-19 hospital outcomes in England before and after vaccine introduction, a cohort study. COVID-hospital-outcomes (GitHub repository). 2022. doi:10.5281/zenodo.6856530

\end{enumerate}

\newpage

\subsection*{Acknowledgements}

We gratefully acknowledge all the clinicians, data reporters and individuals whose data were used in this study, as well as all UK Health Security Agency colleagues involved in the COVID-19 response. We thank Shaun Seaman and Tommy Nyberg of the MRC Biostatistics Unit, Cambridge for their advice on epidemic phase bias, Kevin Fong and Tristan Caulfield at University College London Hospitals NHS Foundation Trust for discussion of the hospital load measure, James Stimson and members of the UKHSA Joint Modelling Team for the discussion and support of these analyses, and members of the UKHSA Immunisation Division for support with data preparation and linkage.

This research is funded by the Medical Research Council (Unit programme number MC\_UU\_00002/11, DDA, AMP); a grant from the MRC UKRI/DHSC NIHR COVID-19 rapid response call (grant ref: MC\_PC\_19074, DDA, AMP); and the NIHR Health Protection Research Unit in Behavioural Science and Evaluation (DDA, AMP). This research is also funded by the Department of Health and Social Care using UK Aid funding, managed by the NIHR (grant number PR-OD-1017-20006, DDA, AMP). The funders had no influence on the methods, interpretation of results or decision to submit.

\subsection*{Authors’ contributions statement}

PDK, PB, AMP, and DDA conceived the research study. AC, RH, SE, and SM undertook data collection and dataset generation. PDK and AMP drafted the manuscript and formatted and verified the datasets. PDK carried out the analyses. AC, PB, SE, SM, RH, and DDA provided expert advice and critical review of the manuscript prior to submission. The corresponding author attests that all listed authors meet authorship criteria and that no others meeting the criteria have been omitted.

\subsection*{Competing interests statement}

The authors declare no competing interests.

\cleardoublepage

\section*{Tables}

\footnotesize

\begin{longtable}{@{}llllll@{}}
\caption*{Table 1: Characteristics of the study population compared with all people hospital-onset COVID-19 in England and all people with PCR-confirmed community-acquired COVID-19 in England.}
\label{tab:my-table}\\
\toprule
\multicolumn{2}{l}{\textbf{Characteristic}} &
  \textbf{\begin{tabular}[c]{@{}l@{}}Study population \\ (hospitalised for \\ COVID-19 in \\ England)\end{tabular}} &
  \multicolumn{2}{l}{\textbf{\begin{tabular}[c]{@{}l@{}}All people with \\ hospital-onset \\ COVID-19 in \\ England\end{tabular}}} &
  \textbf{\begin{tabular}[c]{@{}l@{}}All people with \\ PCR-confirmed \\ community-\\ acquired COVID-\\ 19 in England\end{tabular}} \\* \midrule
\endfirsthead
\multicolumn{6}{c}%
{{\bfseries Table 1 continued from previous page}} \\
\endhead
\bottomrule
\endfoot
\endlastfoot
           &                                               & n   (\%)         & \multicolumn{2}{l}{n   (\%)}         & n   (\%)           \\
\multicolumn{2}{l}{\textbf{Total}}                         & 259,727 (100\%)  & \multicolumn{2}{l}{208,851 (100\%)}  & 6,616,231 (100\%)  \\
\multicolumn{2}{l}{\textbf{Age}}                           &                  & \multicolumn{2}{l}{}                 &                    \\
           & 0-14                                          & 6650 (2.6\%)     & \multicolumn{2}{l}{2632 (1.3\%)}     & 892,640 (13.5\%)   \\
           & 15-24                                         & 8972 (3.5\%)     & \multicolumn{2}{l}{4417 (2.1\%)}     & 1,265,595 (19.1\%) \\
           & 35-44                                         & 44,094 (17.0\%)  & \multicolumn{2}{l}{13728 (6.6\%)}    & 2,242,751 (33.9\%) \\
           & 45-64                                         & 74,258 (28.6\%)  & \multicolumn{2}{l}{33,166 (15.9\%)}  & 1,559,808 (23.6\%) \\
           & 65-74                                         & 42,307 (16.3\%)  & \multicolumn{2}{l}{36,466 (17.5\%)}  & 301,356 (4.6\%)    \\
           & 75-84                                         & 47,783 (18.4\%)  & \multicolumn{2}{l}{59,841 (28.7\%)}  & 198,201 (3.0\%)    \\
           & 85+                                           & 35,663 (13.7\%)  & \multicolumn{2}{l}{58,601 (28.1\%)}  & 155,880 (2.4\%)    \\
\multicolumn{2}{l}{\textbf{Sex}}                           &                  & \multicolumn{2}{l}{}                 &                    \\
           & Male                                          & 135,419 (52.1\%) & \multicolumn{2}{l}{108,518 (52.0\%)} & 3,151,711 (47.6\%) \\
           & Female                                        & 124,308 (47.9\%) & \multicolumn{2}{l}{100,333 (48.0\%)} & 3,464,520 (52.4\%) \\
\multicolumn{2}{l}{\textbf{Ethnicity}}                     &                  & \multicolumn{2}{l}{}                 &                    \\
           & White                                         & 195,496 (75.3\%) & \multicolumn{2}{l}{185,078 (88.6\%)} & 5,100,116 (77.1\%) \\
           & Asian                                         & 32,191 (12.4\%)  & \multicolumn{2}{l}{9930 (4.8\%)}     & 740,194 (11.2\%)   \\
           & Black                                         & 15,394 (5.9\%)   & \multicolumn{2}{l}{6376 (3.1\%)}     & 278,773 (4.2\%)    \\
           & Mixed/Other/Unknown                           & 16,646 (6.4\%)   & \multicolumn{2}{l}{7467 (3.6\%)}     & 497,148 (7.5\%)    \\
\multicolumn{2}{l}{\textbf{Region of residence}}           &                  & \multicolumn{2}{l}{}                 &                    \\
           & London                                        & 46,854 (18.0\%)  & \multicolumn{2}{l}{28,976 (13.9\%)}  & 1,077,599 (16.3\%) \\
           & East Midlands                                 & 22,571 (8.7\%)   & \multicolumn{2}{l}{18,721 (9.0\%)}   & 586,513 (8.9\%)    \\
           & East of England                               & 26,323 (10.1\%)  & \multicolumn{2}{l}{22,696 (10.9\%)}  & 678,869 (10.3\%)   \\
           & North East                                    & 15,131 (5.8\%)   & \multicolumn{2}{l}{9403 (4.5\%)}     & 376,600 (5.7\%)    \\
           & North West                                    & 43,017 (16.6\%)  & \multicolumn{2}{l}{42,130 (20.2\%)}  & 1,056,012 (16.0\%) \\
           & South East                                    & 32,740 (12.6\%)  & \multicolumn{2}{l}{28,785 (13.8\%)}  & 887,116 (13.4\%)   \\
           & South West                                    & 17,391 (6.7\%)   & \multicolumn{2}{l}{12,854 (6.2\%)}   & 496,411 (7.5\%)    \\
           & West Midlands                                 & 29,808 (11.5\%)  & \multicolumn{2}{l}{24,513 (11.7\%)}  & 733,250 (11.1\%)   \\
           & Yorkshire and Humber                          & 25,892 (10.0\%)  & \multicolumn{2}{l}{20,773 (9.9\%)}   & 723,861 (10.9\%)   \\
\multicolumn{2}{l}{\textbf{Index of multiple deprivation}} & \multicolumn{2}{l}{}                  & \multicolumn{2}{l}{}                 \\
           & 1st quintile (most deprived)                  & 73,100 (28.1\%)  & \multicolumn{2}{l}{51,176 (24.5\%)}  & 1,564,464 (23.6\%) \\
           & 2nd quintile                                  & 59,754 (23.0\%)  & \multicolumn{2}{l}{44,886 (21.5\%)}  & 1,440,534 (21.8\%) \\
           & 3rd quintile                                  & 48,458 (18.7\%)  & \multicolumn{2}{l}{40,595 (19.4\%)}  & 1,287,972 (19.5\%) \\
           & 4th quintile                                  & 42,608 (16.4\%)  & \multicolumn{2}{l}{38,396 (18.4\%)}  & 1,209,869 (18.3\%) \\
           & 5th quintile (least deprived)                 & 35,807 (13.8\%)  & \multicolumn{2}{l}{33,798 (16.2\%)}  & 1,113,392 (16.8\%) \\
\multicolumn{2}{l}{\textbf{Month of hospital admission}}   & \multicolumn{2}{l}{}                  & \multicolumn{2}{l}{}                 \\
           & Mar-20                                        & 12,408 (4.8\%)   & \multicolumn{2}{l}{17,564 (8.4\%)}   & 31,598 (0.5\%)     \\
           & Apr-20                                        & 25,867 (10.0\%)  & \multicolumn{2}{l}{29,525 (14.1\%)}  & 111,629 (1.7\%)    \\
           & May-20                                        & 7475 (2.9\%)     & \multicolumn{2}{l}{10,170 (4.9\%)}   & 66,563 (1.0\%)     \\
           & Jun-20                                        & 2698 (1.0\%)     & \multicolumn{2}{l}{3740 (1.8\%)}     & 25,007 (0.4\%)     \\
           & Jul-20                                        & 1077 (0.4\%)     & \multicolumn{2}{l}{1232 (0.6\%)}     & 18,905 (0.3\%)     \\
           & Aug-20                                        & 911 (0.4\%)      & \multicolumn{2}{l}{514 (0.2\%)}      & 29,130 (0.4\%)     \\
           & Sep-20                                        & 3945 (1.5\%)     & \multicolumn{2}{l}{2092 (1.0\%)}     & 124,294 (1.9\%)    \\
           & Oct-20                                        & 15,235 (5.9\%)   & \multicolumn{2}{l}{14,004 (6.7\%)}   & 474,083 (7.2\%)    \\
           & Nov-20                                        & 23,218 (8.9\%)   & \multicolumn{2}{l}{22,905 (11.0\%)}  & 518,220 (7.8\%)    \\
           & Dec-20                                        & 31,468 (12.1\%)  & \multicolumn{2}{l}{31,564 (15.1\%)}  & 852,653 (12.9\%)   \\
           & Jan-21                                        & 60,389 (23.3\%)  & \multicolumn{2}{l}{40,122 (19.2\%)}  & 1,068,457 (16.1\%) \\
           & Feb-21                                        & 19,503 (7.5\%)   & \multicolumn{2}{l}{12,959 (6.2\%)}   & 289,488 (4.4\%)    \\
           & Mar-21                                        & 5809 (2.2\%)     & \multicolumn{2}{l}{3750 (1.8\%)}     & 134,516 (2.0\%)    \\
           & Apr-21                                        & 1929 (0.7\%)     & \multicolumn{2}{l}{1110 (0.5\%)}     & 58,209 (0.9\%)     \\
           & May-21                                        & 1370 (0.5\%)     & \multicolumn{2}{l}{477 (0.2\%)}      & 61,016 (0.9\%)     \\
           & Jun-21                                        & 3756 (1.4\%)     & \multicolumn{2}{l}{1325 (0.6\%)}     & 293,512 (4.4\%)    \\
           & Jul-21                                        & 13,984 (5.4\%)   & \multicolumn{2}{l}{4377 (2.1\%)}     & 926,889 (14.0\%)   \\
           & Aug-21                                        & 15,578 (6.0\%)   & \multicolumn{2}{l}{6394 (3.1\%)}     & 775,051 (11.7\%)   \\
           & Sep-21                                        & 13,107 (5.0\%)   & \multicolumn{2}{l}{5027 (2.4\%)}     & 757,011 (11.4\%)   \\
\multicolumn{2}{l}{\textbf{Hospital outcome}}              &                  & \multicolumn{2}{l}{}                 &                    \\
           & Death                                         & 51,948 (20.0\%)  & \multicolumn{2}{l}{69,243 (33.2\%)}  & N/A                \\
           & Discharge                                     & 191,663 (73.8\%) & \multicolumn{2}{l}{107,815 (51.6\%)} & N/A                \\
           & Right-censored in hospital                    & 16,116 (6.2\%)   & \multicolumn{2}{l}{31,793 (15.2\%)}  & N/A                \\
\multicolumn{6}{l}{\textbf{Median length of stay following admission/positive test (days)}}                                               \\
           & Death                                         & 8 days           & \multicolumn{2}{l}{11 days}          & N/A                \\
           & Discharge                                     & 5 days           & \multicolumn{2}{l}{13 days}          & N/A                \\
\multicolumn{6}{l}{\textbf{Vaccination status at date of admission (for admissions occurring January and July 2021)}}                     \\
           & Unvaccinated                                  & 97,441 (72.0\%)  & \multicolumn{2}{l}{N/A}              & N/A                \\
           & \textless{}21 days after first dose           & 10,774 (8.0\%)   & \multicolumn{2}{l}{N/A}              & N/A                \\
           & $\geq$21 days after first dose                     & 7885 (5.8\%)     & \multicolumn{2}{l}{N/A}              & N/A                \\
           & $\geq$14 days after second dose                    & 19,325 (14.3\%)  & \multicolumn{2}{l}{N/A}              & N/A                \\
\multicolumn{2}{l}{\textbf{Charlson comorbidity index}}    & \multicolumn{2}{l}{}                  & \multicolumn{2}{l}{}                 \\
           & 0                                             & 92,753 (38.8\%)  & \multicolumn{2}{l}{N/A}              & N/A                \\
           & 1-2                                           & 93,436 (39.1\%)  & \multicolumn{2}{l}{N/A}              & N/A                \\
           & 3-4                                           & 35,527 (14.9\%)  & \multicolumn{2}{l}{N/A}              & N/A                \\
           & 5+                                            & 17,190 (7.2\%)   & \multicolumn{2}{l}{N/A}              & N/A                \\
\multicolumn{6}{l}{\textbf{Hospital load at time of admission (as proportion of busiest week)}}                                           \\
           & 0-20\%                                        & 61,406 (23.6\%)  & \multicolumn{2}{l}{N/A}              & N/A                \\
           & 20-40\%                                       & 64,722 (24.9\%)  & \multicolumn{2}{l}{N/A}              & N/A                \\
           & 40-60\%                                       & 49,529 (19.1\%)  & \multicolumn{2}{l}{N/A}              & N/A                \\
           & 60-80\%                                       & 40,385 (15.5\%)  & \multicolumn{2}{l}{N/A}              & N/A                \\
           & 80-90\%                                       & 20,228 (7.8\%)   & \multicolumn{2}{l}{N/A}              & N/A                \\
           & 90-100\%                                      & 23,457 (9.0\%)   & \multicolumn{2}{l}{N/A}              & N/A                \\
\multicolumn{2}{l}{\textbf{Route of admission}}            &                  & \multicolumn{2}{l}{}                 &                    \\
           & Via emergency ward                            & 204,151 (78.6\%) & \multicolumn{2}{l}{N/A}              & N/A                \\
           & Directly to hospital                          & 55,576 (21.4\%)  & \multicolumn{2}{l}{N/A}              & N/A                \\* \bottomrule
\end{longtable}

\begin{longtable}{@{}lllll@{}}
\caption*{Table 2: Hospitalised fatality risk by vaccine status at hospital admission and age group. Figures in brackets represent 95\% confidence intervals. Estimates are replaced with a dash (-) where insufficient information was available.}
\label{tab:my-table}\\
\toprule
\textbf{\begin{tabular}[c]{@{}l@{}}Age \\ group\end{tabular}} &
  \textbf{Unvaccinated} &
  \textbf{\begin{tabular}[c]{@{}l@{}}\textless{}21   days after \\ first dose\end{tabular}} &
  \textbf{\begin{tabular}[c]{@{}l@{}}$\geq$21   days after \\ first dose\end{tabular}} &
  \textbf{\begin{tabular}[c]{@{}l@{}}$\geq$14   days after \\ second dose\end{tabular}} \\* \midrule
\endfirsthead
\multicolumn{5}{c}%
{{\bfseries Table 2 continued from previous page}} \\
\endhead
\bottomrule
\endfoot
\endlastfoot
{[}0,15)     & 0.2\% (0.1 - 0.4\%)    & -                      & -                      & -                      \\
{[}15,25)    & 0.3\% (0.2 - 0.5\%)    & 1.3\% (0.3 - 5.3\%)    & -                      & -                      \\
{[}25,45)    & 1.7\% (1.5 - 1.9\%)    & 2.1\% (1.3 - 3.5\%)    & 1.2\% (0.7 - 2.2\%)    & -                      \\
{[}45,65)    & 10.1\% (9.8 - 10.5\%)  & 8.7\% (7.5 - 10.1\%)   & 6.1\% (4.9 - 7.7\%)    & 7.5\% (6.2 - 9.0\%)    \\
{[}65,75)    & 25.3\% (24.5 - 26.0\%) & 22.5\% (20.5 - 24.7\%) & 18.2\% (15.6 - 21.1\%) & 14.9\% (12.9 - 17.2\%) \\
{[}75,85)    & 38.6\% (37.7 - 39.6\%) & 34.5\% (32.8 - 36.2\%) & 26.2\% (24.1 - 28.5\%) & 22.5\% (20.4 - 24.8\%) \\
{[}85,Inf{]} & 49.9\% (48.7 - 51.1\%) & 47.0\% (45.2 - 48.9\%) & 37.7\% (35.4 - 40.2\%) & 32.0\% (29.1 - 35.2\%) \\* \bottomrule
\end{longtable}

\normalsize

\cleardoublepage

\section*{Figures}

\begin{figure}[htbp!]
\centering
\caption*{\textbf{Figure 1: Observed number of individuals hospitalised with COVID-19, by week of admission.}}
\includegraphics[scale=0.45]{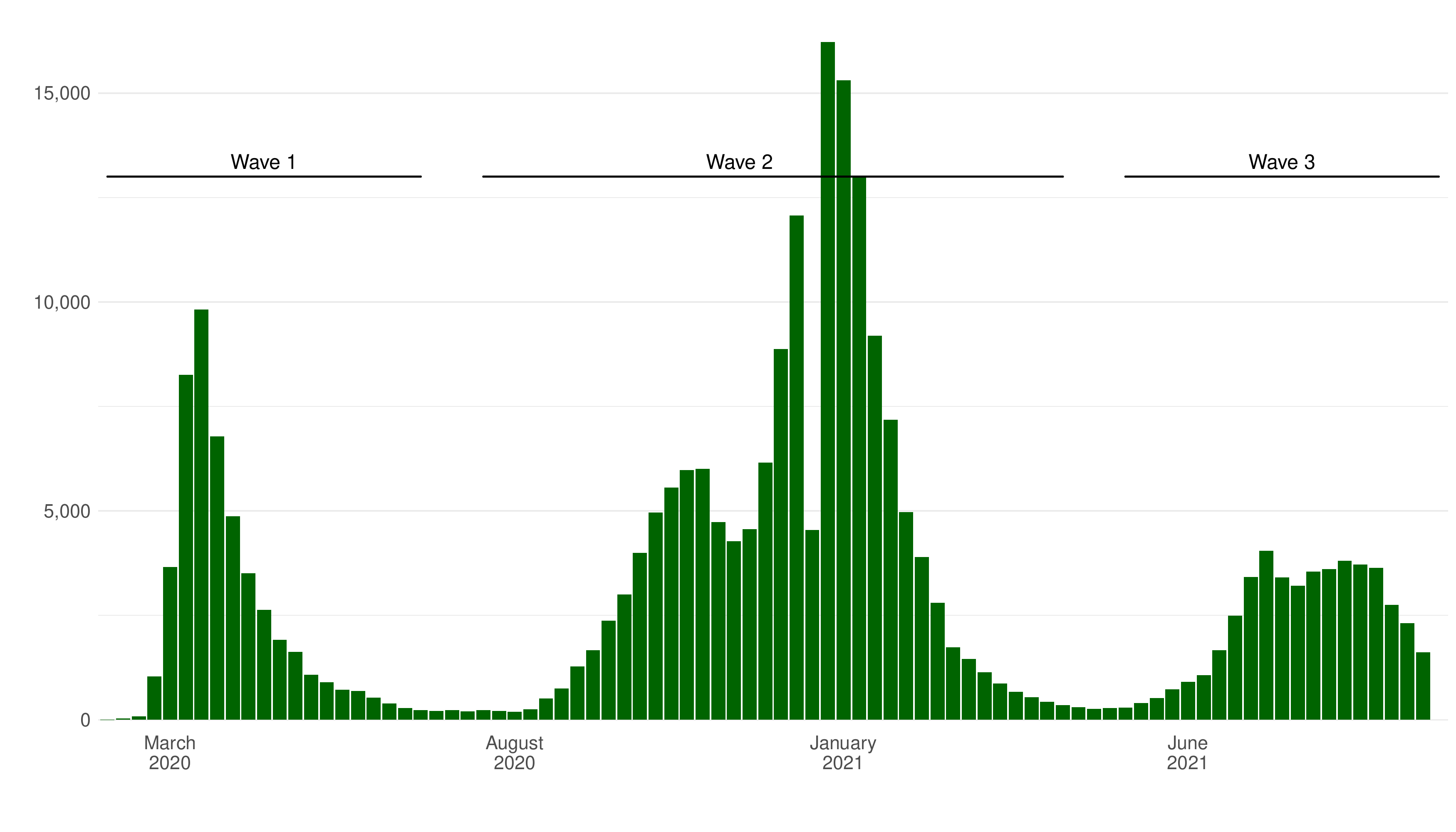}
\caption*{March 2020 to September 2021. Annotations shown for wave 1, wave 2, and wave 3. $n=259,727$ individuals.}
\end{figure}

\begin{figure}[htbp!]
\centering
\caption*{\textbf{Figure 2: Vaccination status of hospitalised individuals, by month of admission and age group.}}
\includegraphics[scale=0.45]{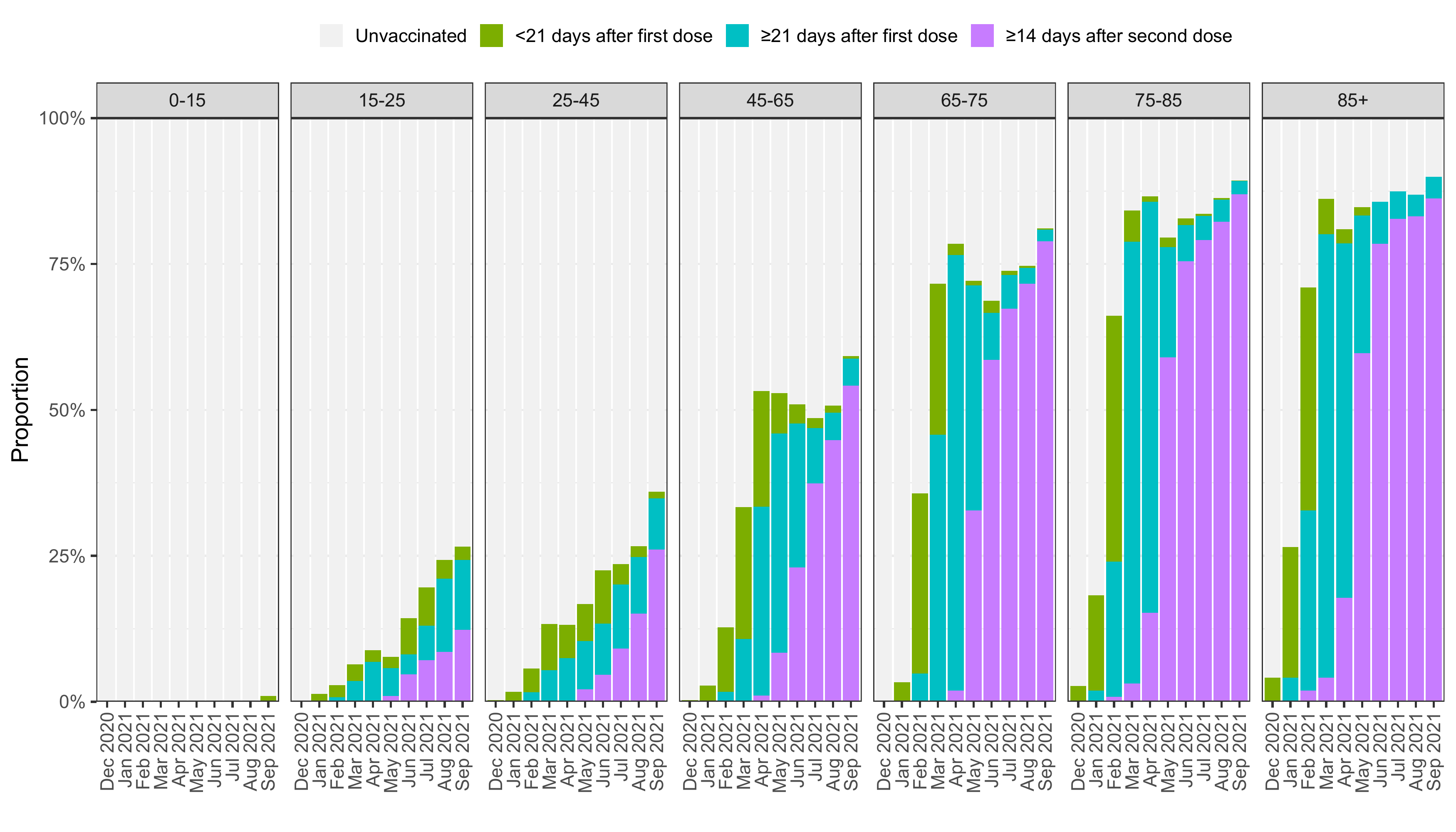}
\caption*{December 2020 to September 2021. $n=166,893$ individuals with vaccination status reported.}
\end{figure}

\begin{figure}[htbp!]
\centering
\caption*{\textbf{Figure 3: Hospitalised fatality risk (panel a) and median length of stay (panel b) by month of admission.}}
\includegraphics[scale=0.55]{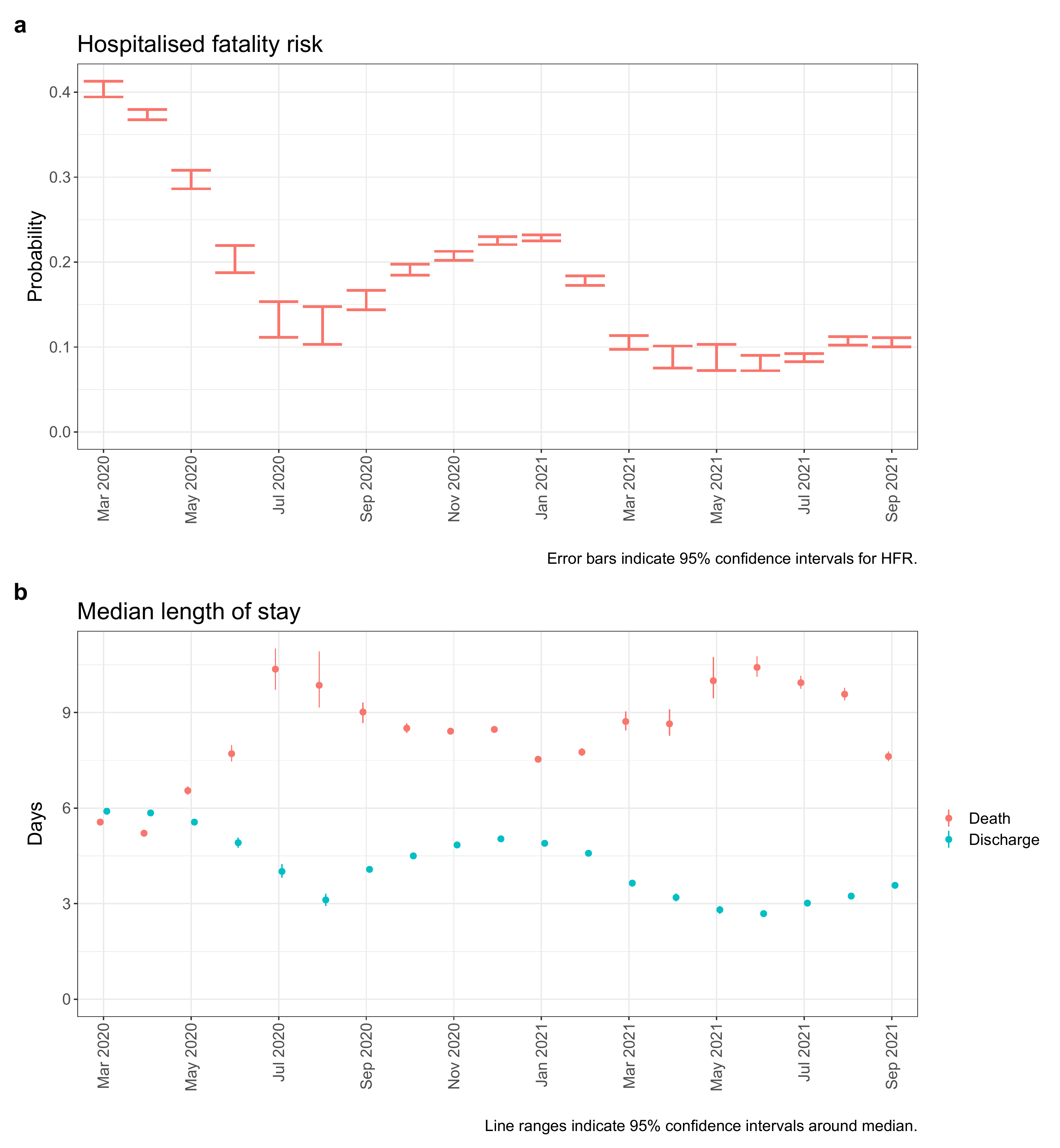}
\caption*{March 2020 to September 2021. Unadjusted for other covariates. $n=259,727$ individuals. Error bars are 95\% confidence intervals.}
\end{figure}

\begin{figure}[htbp!]
\centering
\caption*{\textbf{Figure 4: Hospitalised fatality risk (panel a) and median length of stay (panel b) by month of admission and age group.}}
\includegraphics[scale=0.55]{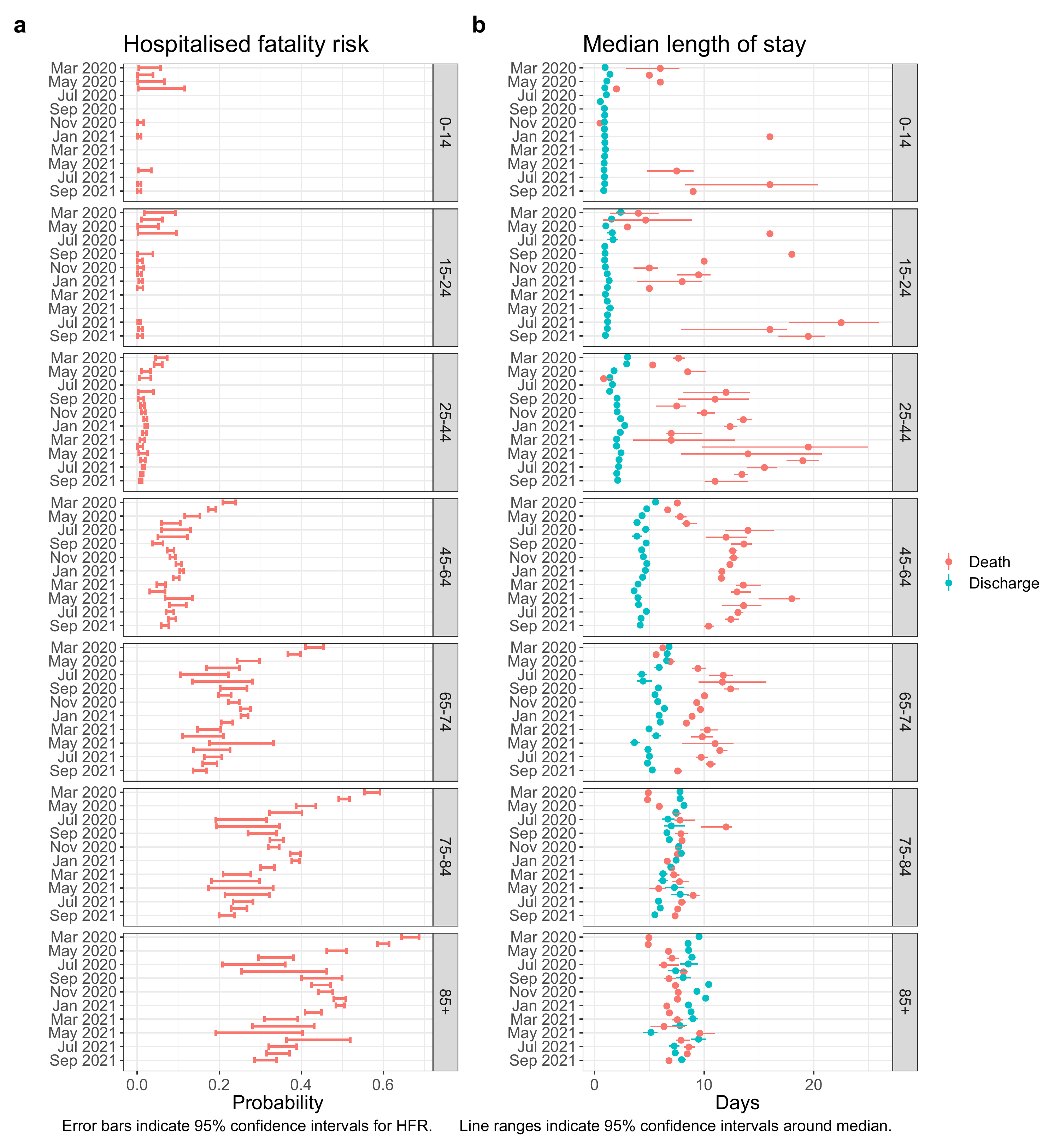}
\caption*{March 2020 to September 2021. Unadjusted for other covariates. $n=259,727$ individuals with age reported. Error bars are 95\% confidence intervals.}
\end{figure}

\begin{figure}[htbp!]
\centering
\caption*{\textbf{Figure 5: Hospitalised fatality sub-distribution hazard ratio by month of hospital admission.}}
\includegraphics[scale=0.55]{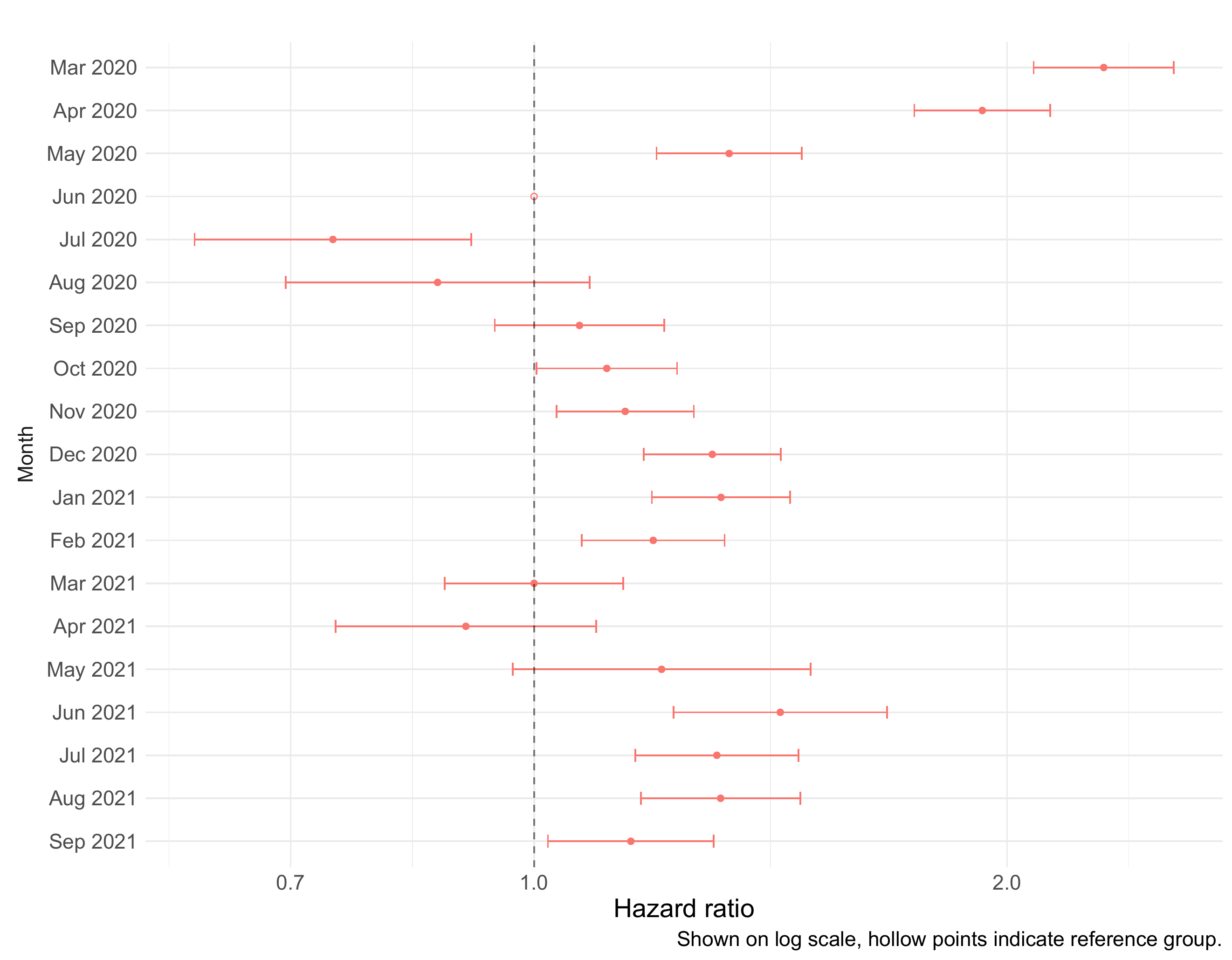}
\caption*{March 2020 to September 2021. Model includes stratification on age group, region of residence, and vaccination status, and regression adjustment (main effects) on month of hospital admission, sex, ethnicity, IMD quintile, hospital load, and CCI. Reference group: June 2020. $n=238,897$ individuals with necessary information reported. Figure shows point estimate of hazard ratio with 95\% confidence intervals.}
\end{figure}

\begin{figure}[htbp!]
\centering
\caption*{\textbf{Figure 6: Hospitalised fatality sub-distribution hazard ratio by vaccine status.}}
\includegraphics[scale=0.55]{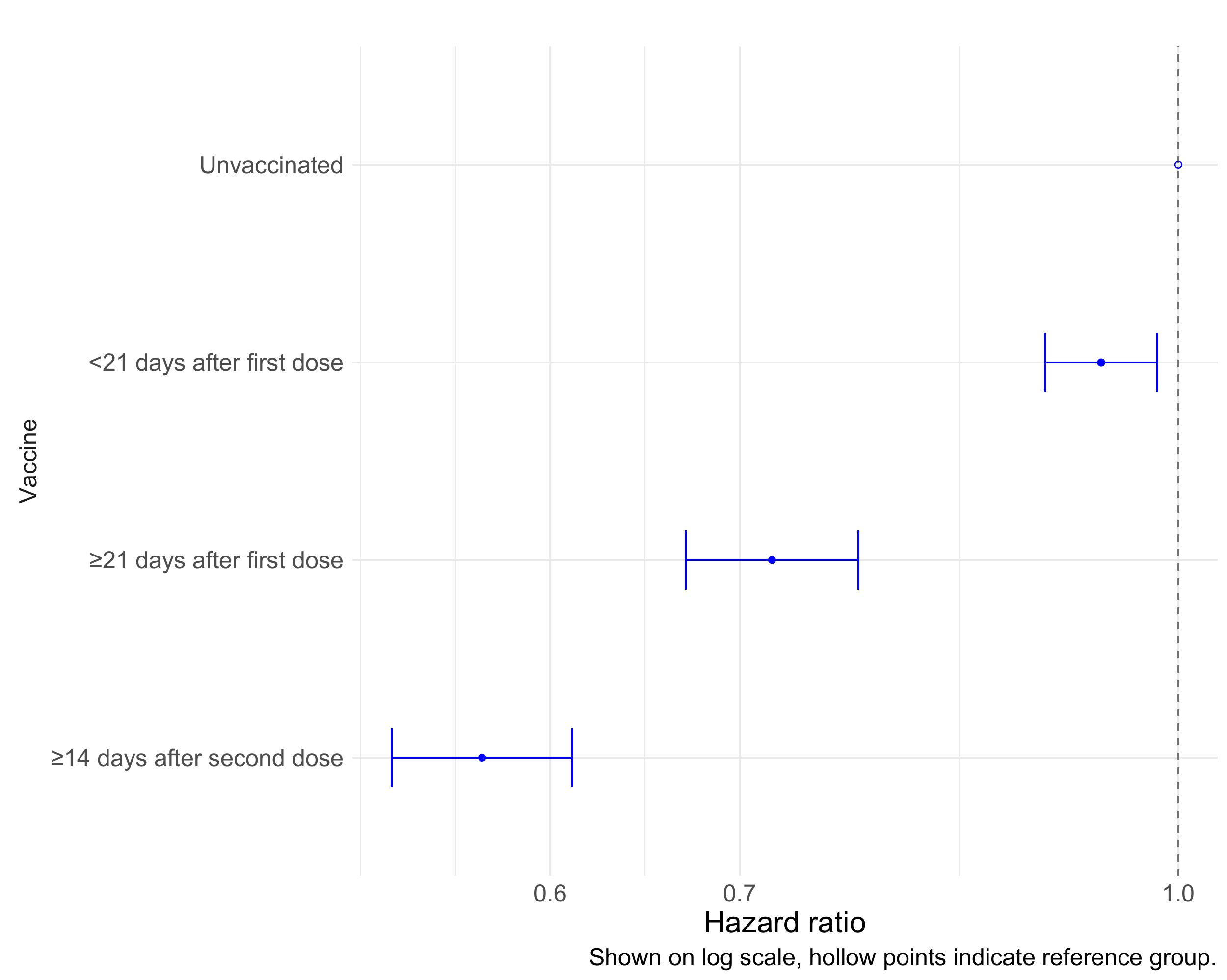}
\caption*{January 2021 to September 2021. Model includes stratification on age group, region of residence, and month of hospital admission, and regression adjustment (main effects) on vaccine status, sex, ethnicity, IMD quintile, hospital load, and CCI. Reference group: Unvaccinated. $n=126,679$ individuals with necessary information reported. Figure shows point estimate of hazard ratio with 95\% confidence intervals.}
\end{figure}

\newpage

\section*{Supplementary information}

\section*{Aalen-Johansen estimator}

If the competing risks of events are not independent, and the magnitude of the competing risk is large, the assumptions of conventional survival analysis methods such as Kaplan-Meier and Cox proportional hazards regression may provide biased estimates of risk, or of covariate effects on the rate of events, respectively [1]. In the case of mortality within 90 days of hospitalisation with COVID-19, the competing risk of discharge may well be large, e.g. for individuals discharged to palliative care.

The Aalen-Johansen estimator is the standard non-parametric estimate of the cumulative incidence function for competing risk [2], also described as the matrix version of the Kaplan-Meier estimator. 

Let the transition hazard from state $i \in S$ to state $j \in S$, $i \neq j$ be defined as:
$$
\alpha_{i,j}(t)dt=P(X_{t+dt}=j\ |\ X_t=i)
$$

and let $\mathbf{A}(t)$ be the matrix of cumulative transition hazards:
\begin{gather*}
A_{i,j}(t)=\int_{0}^{t}{\alpha_{i,j}(u)du} \\ 
A_{i,i}(t)\ =\ -\sum_{j\neq i}{A_{i,j}(t)}
\end{gather*}

then, defining as $P_{i,j}(s,t)=P(X_t=j\ |\ X_s=i),$ for $i,j \in S$, $s \le t$ the probability that an individual in state $s$ at time $i$ will be in state $t$ at time $j$, the Aalen-Johansen estimator is given by the matrix of transition probabilities:
$$
\mathbf{P}(s,t)=\prod_{s,t}{(\mathbf{I}+d\mathbf{A}(u))}
$$

The cumulative incidence function is a special case of the Aalen-Johansen estimate where:
$$
C_k(t)=\ \int_{0}^{t}{\alpha_k(u)S(u)du}
$$
$\alpha_k$ is the incidence function for outcome $k$, and $S(u)$ is the overall survival curve.

\section*{Fine-Gray proportional hazards regression}

The Fine-Gray model estimates the hazard of a competing event (so-termed the sub-distribution hazard) among the risk set of those yet to experience an event of the type of interest by time $t$ [3]. The risk set therefore consists of both those who have yet to experience any event and those who have yet to experience the event of interest (e.g. death) but have experienced a competing event (e.g. discharge).

The subdistribution hazard is defined as the instantaneous risk of dying (from a cause $k$) given that the individual has not already died:

$$
h_k(t)=\lim_{\delta \rightarrow 0}{\left\{\frac{P(t \le T<t+\delta t, K=k\ |\ T>t \text{ or } (T \le t \text{ \& } K \neq k))}{\delta t}\right\}}
$$

Fine-Gray regression links the subdistribution hazard to the Aalen-Johansen cumulative incidence estimator through the relationship:
$$
h_k(t)=- \frac{d\log{(1-C_k(t))}}{dt}
$$

Covariate effects on the sub-distribution hazard can then be interpreted as covariate effects on the cumulative incidence, or marginal probability, of a competing event (in this case hospitalised fatality).

\section*{Stratification}

Stratified survival analyses enable appropriate adjustment for important confounders, by allowing the baseline hazard to vary across strata [4]. Stratification is a similar principle to matched designs, except rather than a $1:n$ ratio of cases to controls, as many ($a:b$ for $a$ and $b$ both $\geq 1$) cases and controls as possible within each strata are used. 

For the regression on month of admission, stratification was by age group, region of residence and vaccination status, with regression adjustment (main effects) on sex, ethnicity, index of multiple deprivation (IMD) quintile and Charlson comorbidity index (CCI). For the regression on vaccination status, stratification was by age group, region of residence and month of hospital admission, with regression adjustment (main effects) similarly on sex, ethnicity, IMD quintile and CCI.

\section*{Consistency in model estimates}

Supplementary figure 7 demonstrates the high degree of agreement between the Aalen-Johansen and Fine-Gray model estimates.

\section*{Epidemic phase bias}

A form of bias, often ignored in epidemic studies, is the relationship between the time from infection to symptom onset, and an individual’s eventual outcome: e.g. those who go on to die may experience more rapid onset of symptoms following infection. Since estimates must be conditioned on an observed quantity (e.g. symptom onset date or hospital admission date) rather than the unobserved infection date, this relationship can introduce bias into results when an epidemic is in a mode of growth or decline [5]. The resulting bias has been termed “epidemic phase bias” and may result in hazards being over or under-estimated.

To correct for this bias, a time shift of $c$ days should be added to records with the outcome of interest (e.g. mortality), where $c$ is the mean difference in time from infection to symptom onset date between those experiencing the outcome and those not, as proposed by Seaman et al. [5]. As the value of $c$ is typically unknown, sensitivity analysis with differing values of $c$ can be used to assess the susceptibility of results to this bias.

For the sensitivity analysis in this study we shifted the date of symptom onset backwards in time by $c = 0,1,2,3,4$ days for those who died, to mitigate against the effect of more rapid symptom onset for those with more severe illness, where the shift $c$ represents the average difference in time from infection to symptom onset between those patients who died and those who did not. The effect of this shift is shown in Supplementary Figure 8.

\cleardoublepage

\section*{Supplementary figures}

\begin{figure}[htbp!]
\centering
\caption*{\textbf{Supplementary figure 1: Hospitalised fatality risk (panel a) and median length of stay (panel b) by month of admission and sex.}}
\includegraphics[scale=0.55]{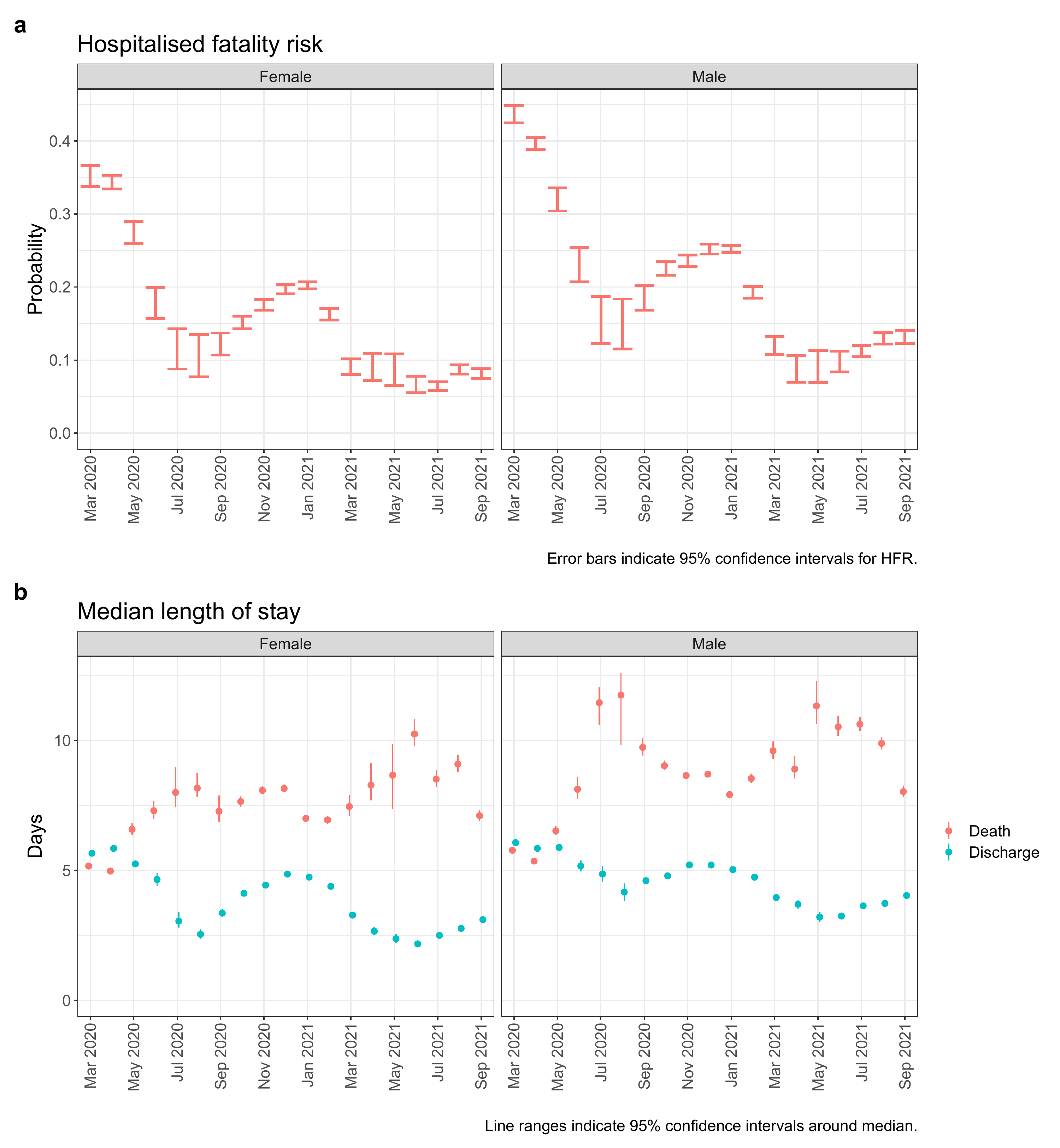}
\caption*{March 2020 to September 2021. Unadjusted for other covariates. $n=259,727$ individuals with sex reported. Error bars are 95\% confidence intervals.}
\end{figure}

\begin{figure}[htbp!]
\centering
\caption*{\textbf{Supplementary figure 2: Hospitalised fatality risk (panel a) and median length of stay (panel b) by month of admission and ethnicity.}}
\includegraphics[scale=0.55]{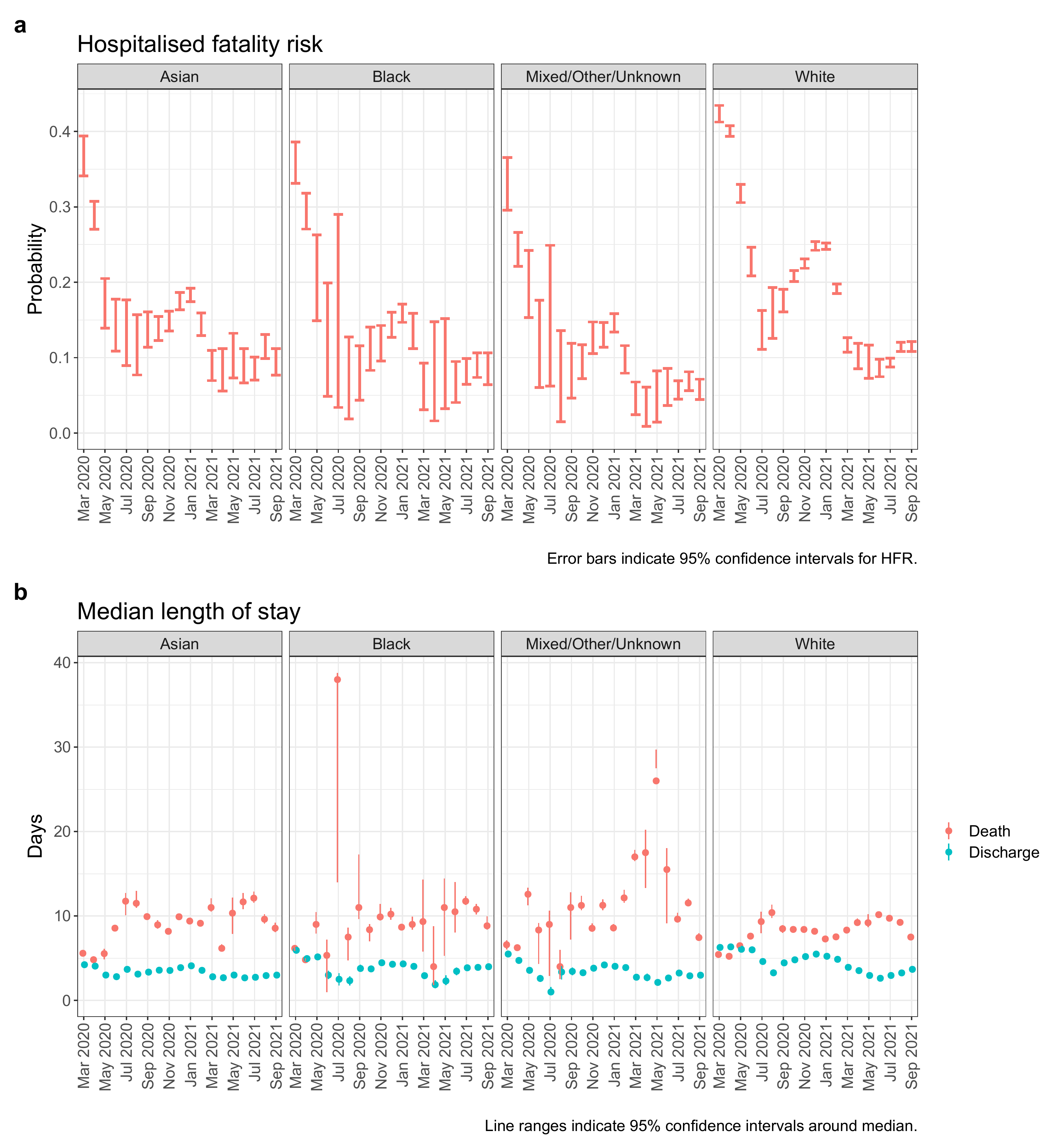}
\caption*{March 2020 to September 2021. Unadjusted for other covariates. $n=259,727$ individuals with ethnicity reported. Error bars are 95\% confidence intervals.}
\end{figure}

\begin{figure}[htbp!]
\centering
\caption*{\textbf{Supplementary figure 3: Hospitalised fatality risk (panel a) and median length of stay (panel b) by month of admission and region of residence.}}
\includegraphics[scale=0.45]{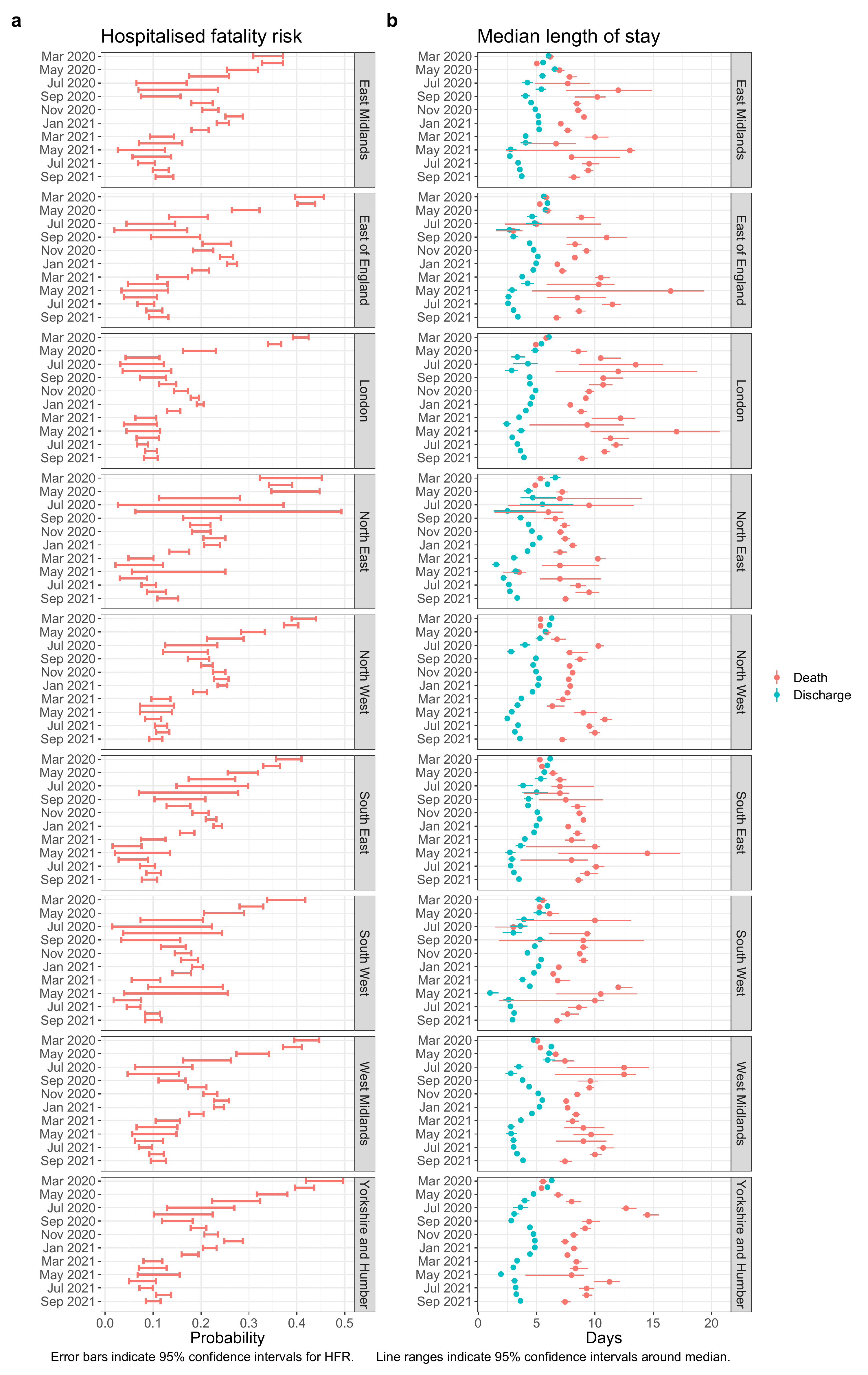}
\caption*{March 2020 to September 2021. Unadjusted for other covariates. $n=259,727$ individuals with region of residence reported. Error bars are 95\% confidence intervals.}
\end{figure}

\begin{figure}[htbp!]
\centering
\caption*{\textbf{Supplementary figure 4: Hospitalised fatality risk (panel a) and median length of stay (panel b) by month of admission and CCI.}}
\includegraphics[scale=0.55]{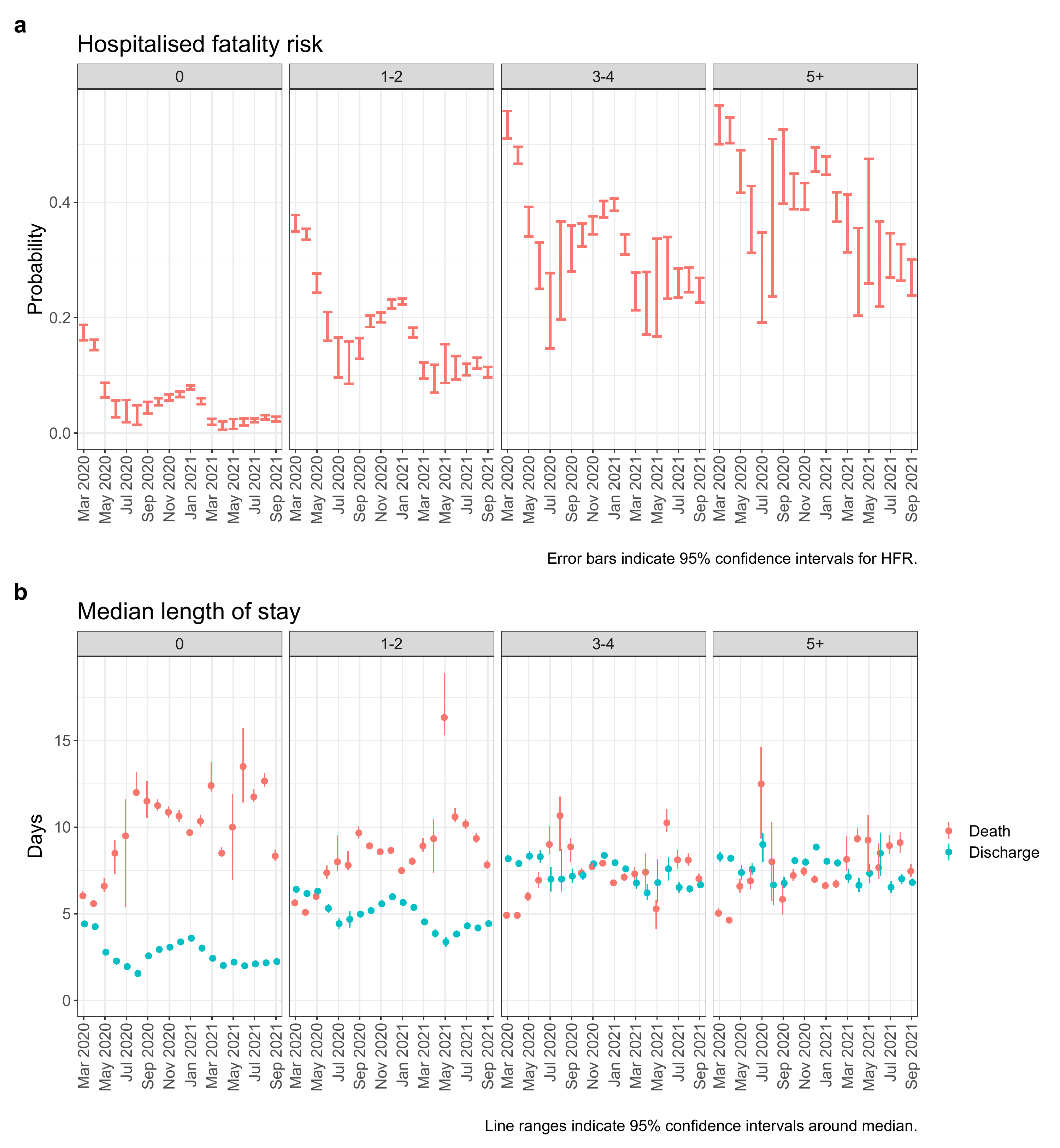}
\caption*{March 2020 to September 2021. Unadjusted for other covariates. $n=238,906$ individuals with comorbidity information reported. Error bars are 95\% confidence intervals.}
\end{figure}

\begin{figure}[htbp!]
\centering
\caption*{\textbf{Supplementary figure 5: Hospitalised fatality risk (panel a) and median length of stay (panel b) by month of admission and measure of hospital load.}}
\includegraphics[scale=0.55]{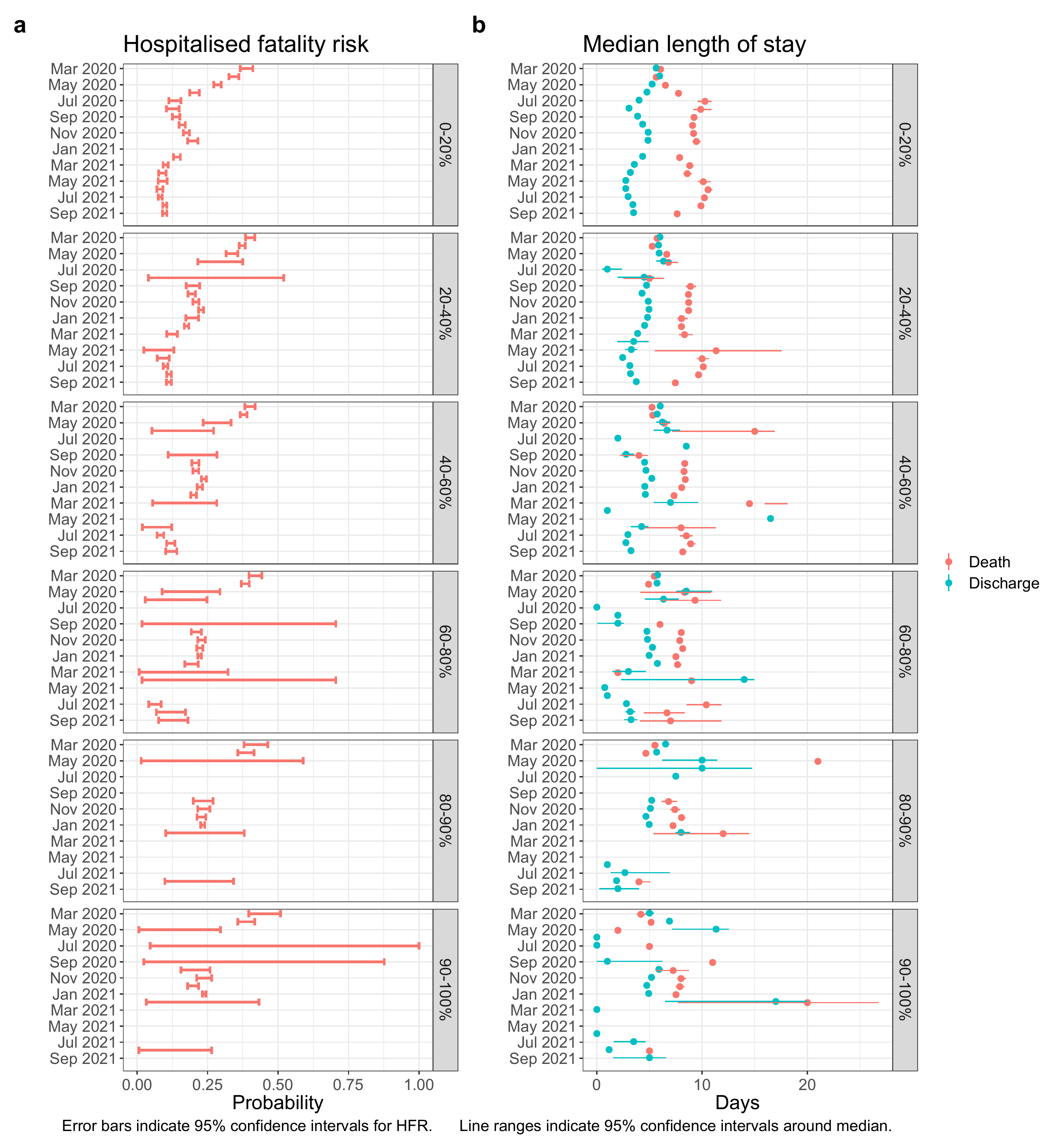}
\caption*{March 2020 to September 2021. Unadjusted for other covariates. $n=259,727$ individuals with hospital admission date reported. Error bars are 95\% confidence intervals.}
\end{figure}

\begin{figure}[htbp!]
\centering
\caption*{\textbf{Supplementary figure 6: Hospitalised fatality sub-distribution hazard ratios for sex, ethnicity, IMD quintile, hospital load and CCI.}}
\includegraphics[scale=0.55]{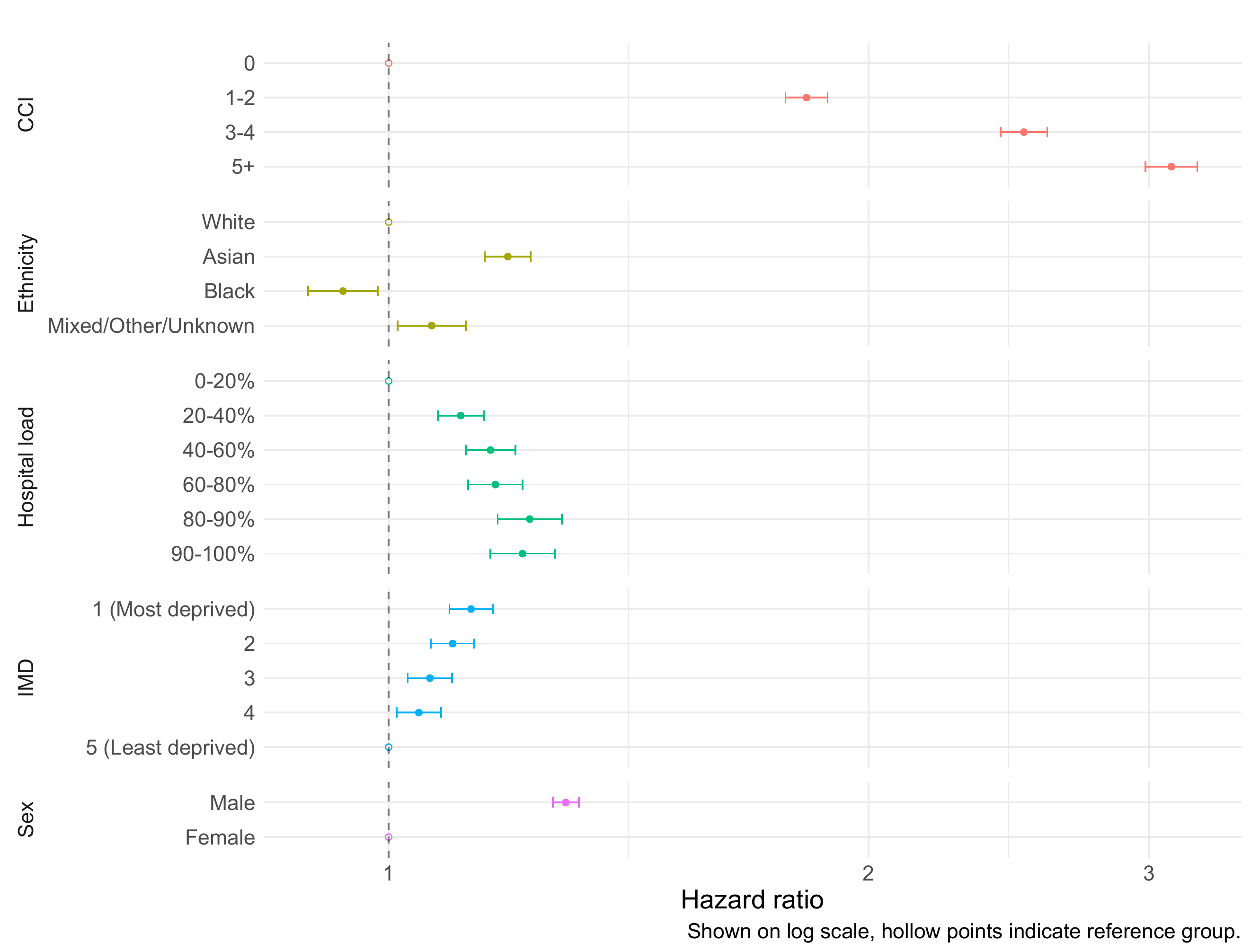}
\caption*{March 2020 to September 2021. Model includes stratification on age group, region of residence, and vaccination status, and regression adjustment (main effects) on month of hospital admission, sex, ethnicity, IMD quintile, hospital load, and CCI. $n=238,897$ individuals with necessary information reported. Figure shows point estimate of hazard ratio with 95\% confidence intervals.}
\end{figure}

\begin{figure}[htbp!]
\centering
\caption*{\textbf{Supplementary figure 7: Aalen-Johansen and Fine-Gray cumulative fatality estimates for first 60 days following hospital admission for selected months.}}
\includegraphics[scale=0.5]{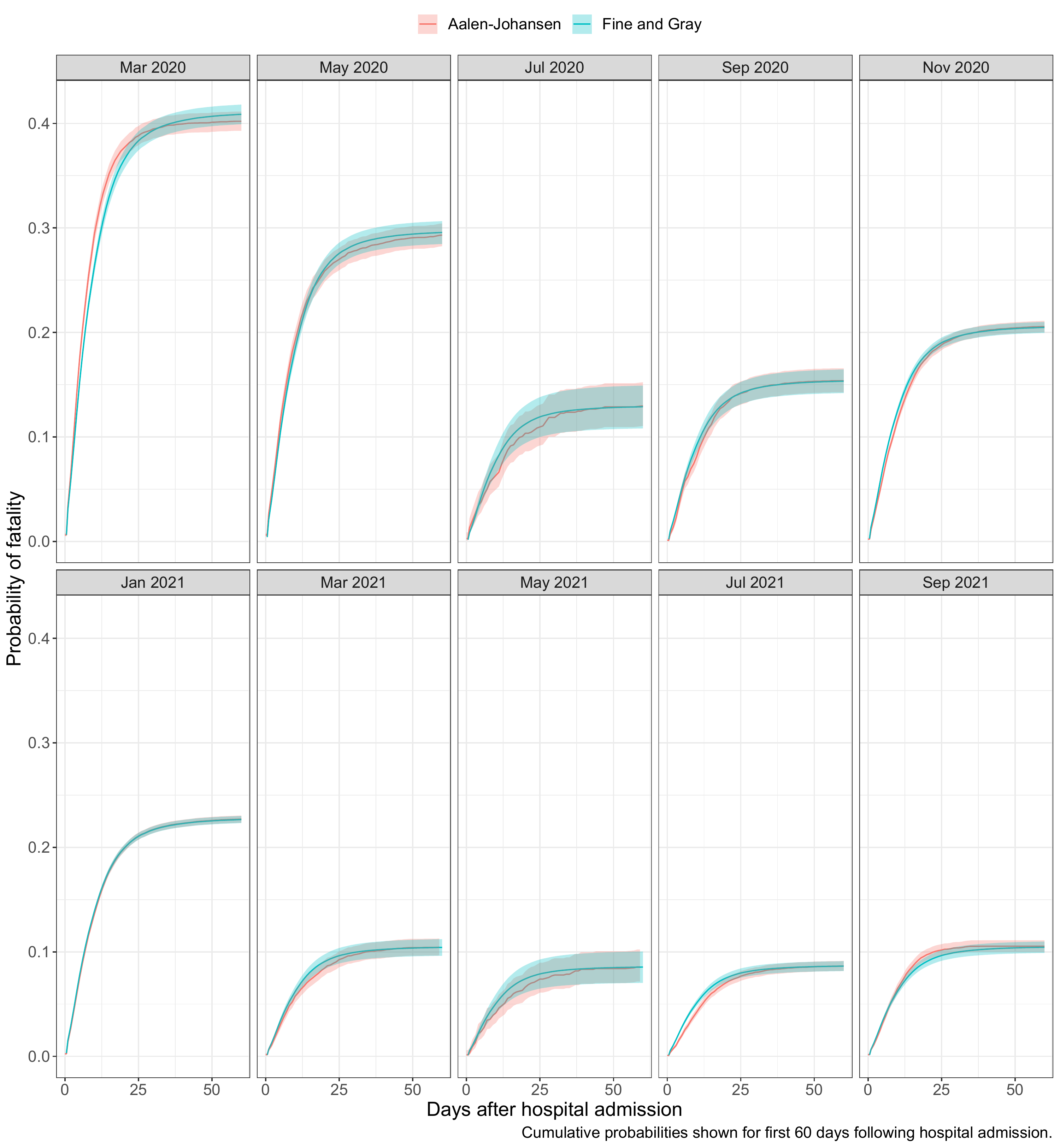}
\caption*{March 2020 to September 2021. $n=259,727$ individuals. Error bands are 95\% confidence intervals. Figure shows high degree of agreement between the two models.}
\end{figure}

\begin{figure}[htbp!]
\centering
\caption*{\textbf{Supplementary figure 8: Hospitalised fatality sub-distribution hazard ratio by month of symptom onset and sensitivity.}}
\includegraphics[scale=0.55]{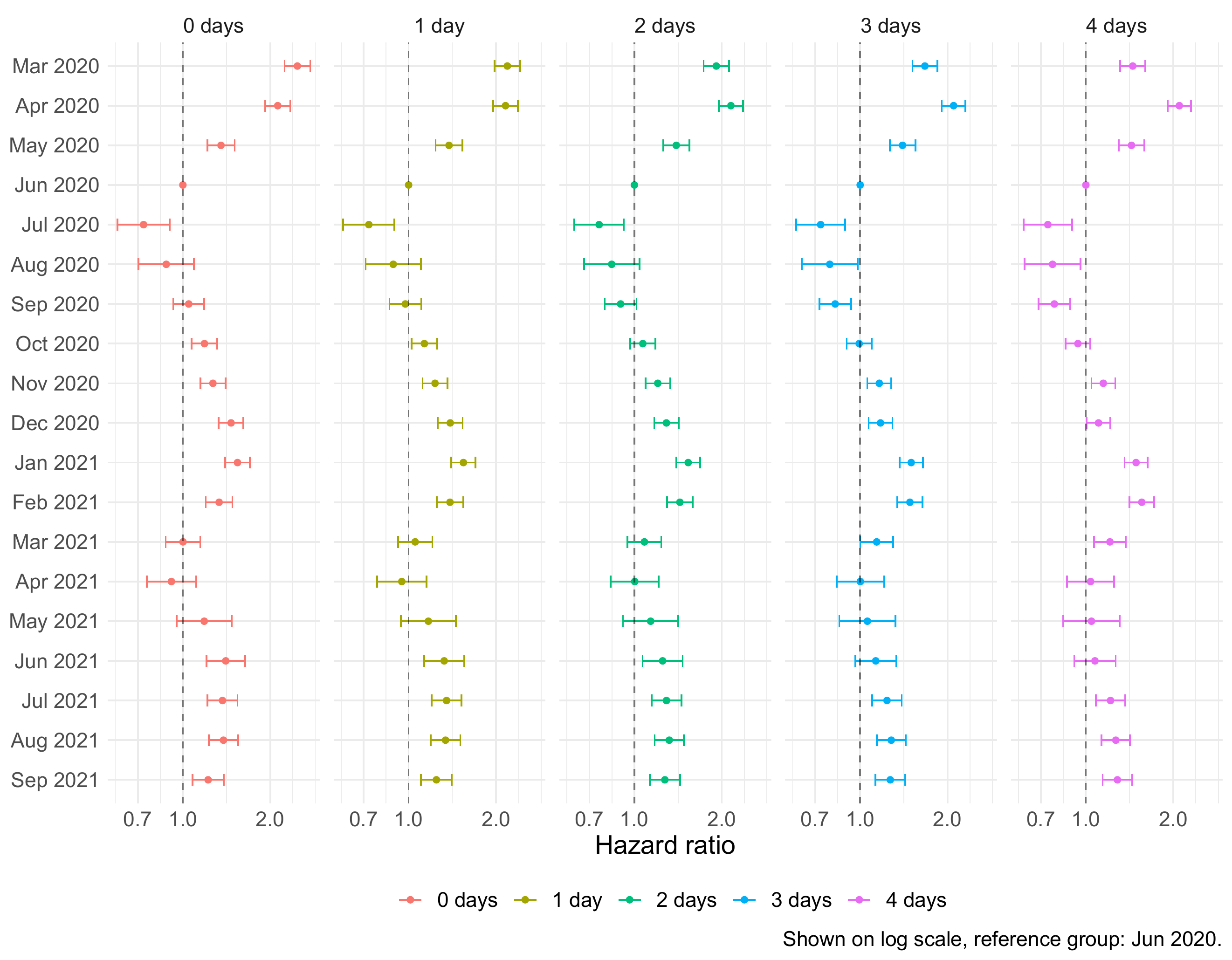}
\caption*{March 2020 to September 2021. Model includes stratification on age group, region of residence, and vaccination status, and regression adjustment (main effects) on month of hospital admission, sex, ethnicity, IMD quintile, hospital load, and CCI. Figure shows results of sensitivity analyses where the date of symptom onset was shifted backwards in time by $c = 0,1,2,3,4$ days for those who died, but not for those who were discharged. Reference group: June 2020. $n=259,727$ individuals with sex reported. Figure shows point estimate of hazard ratio with 95\% confidence intervals.}
\end{figure}

\cleardoublepage

\section*{Supplementary tables}

\begin{longtable}{@{}lllll@{}}
\caption*{Supplementary table 1: Hospital outcomes according to person characteristic. Figures in brackets represent 95\% confidence intervals.}
\label{tab:table1}\\
\toprule
\multicolumn{2}{l}{\textbf{Characteristic}} &
  \textbf{Death} &
  \textbf{Discharge} &
  \textbf{\begin{tabular}[c]{@{}l@{}}Right-censored\\ in hospital\end{tabular}} \\* \midrule
\endfirsthead
\multicolumn{5}{c}%
{{\bfseries Supplementary table 1 continued from previous page}} \\
\endhead
\bottomrule
\endfoot
\endlastfoot
\multicolumn{2}{l}{\textbf{Age}}                           &                &                 &               \\
                    & 0-14                                 & 13 (0.2\%)     & 6272 (94.3\%)   & 365 (5.5\%)   \\
\textbf{}           & 15-24                                & 42 (0.5\%)     & 8509 (94.8\%)   & 421 (4.7\%)   \\
                    & 35-44                                & 735 (1.7\%)    & 40,722 (92.4\%)  & 2637 (6.0\%)    \\
\textbf{}           & 45-64                                & 7310 (9.8\%)   & 61,162 (82.4\%)  & 5786 (7.8\%)  \\
                    & 65-74                                & 10,222 (24.2\%) & 29,369 (69.4\%)  & 2716 (6.4\%)  \\
\textbf{}           & 75-84                                & 17,060 (35.7\%) & 28,133 (58.9\%)  & 2590 (5.4\%)  \\
\textbf{}           & 85+                                  & 16,566 (46.5\%) & 17,496 (49.1\%)  & 1601 (4.5\%)  \\
\bottomrule
\end{longtable}

\footnotesize{*Assessed through existence of matched record in ECDS dataset.}
\normalsize

\cleardoublepage

\begin{longtable}{@{}llll@{}}
\caption*{Supplementary table 2: Hospitalised fatality risk and median length of stay (days) in hospital by month of hospital admission (unadjusted for other covariates). Figures in brackets represent 95\% confidence intervals.}
\label{tab:my-table}\\
\toprule
\textbf{\begin{tabular}[c]{@{}l@{}}Month of\\ hospital admission\end{tabular}} &
  \textbf{Hospitalised fatality risk} &
  \textbf{\begin{tabular}[c]{@{}l@{}}Median LOS\\ prior to death\end{tabular}} &
  \textbf{\begin{tabular}[c]{@{}l@{}}Median LOS\\ prior to discharge\end{tabular}} \\* \midrule
\endfirsthead
\endhead
\bottomrule
\endfoot
\endlastfoot
\textbf{Mar-20} & 40.3\% (39.4 - 41.3\%) & 5.6 (5.5 - 5.6)    & 5.9 (5.8 - 6.0)   \\
\textbf{Apr-20} & 37.4\% (36.7 - 38.0\%)   & 5.2 (5.2 - 5.3)    & 5.8 (5.8 - 5.9) \\
\textbf{May-20} & 29.7\% (28.6 - 30.8\%) & 6.5 (6.4 - 6.7)    & 5.6 (5.5 - 5.7) \\
\textbf{Jun-20} & 20.3\% (18.8 - 21.9\%) & 7.7 (7.5 - 8.0)      & 4.9 (4.8 - 5.1) \\
\textbf{Jul-20} & 13.1\% (11.1 - 15.3\%) & 10.4 (9.7 - 11.0)    & 4.0 (3.8 - 4.2)   \\
\textbf{Aug-20} & 12.3\% (10.3 - 14.8\%) & 9.9 (9.2 - 10.9)   & 3.1 (2.9 - 3.3) \\
\textbf{Sep-20} & 15.5\% (14.4 - 16.7\%) & 9.0 (8.7 - 9.3)      & 4.1 (4.0 - 4.2)   \\
\textbf{Oct-20} & 19.1\% (18.5 - 19.7\%) & 8.5 (8.4 - 8.7)    & 4.5 (4.5 - 4.6) \\
\bottomrule
\end{longtable}

\cleardoublepage

\begingroup\footnotesize
\begin{longtable}{@{}llllll@{}}
\caption*{Supplementary table 3: Hospitalised fatality risk and median length of stay (days) in hospital prior to death or discharge by month of hospital admission and age group (unadjusted for other covariates). Estimates are replaced with a dash (-) where insufficient information was available. Figures in brackets represent 95\% confidence intervals.}
\label{tab:my-table}\\
\toprule
\multicolumn{2}{l}{\textbf{\begin{tabular}[c]{@{}l@{}}Month of hospital \\ admission\end{tabular}}} & \textbf{0-14}        & \textbf{15-24}      & \textbf{25-44}      & \textbf{45-64}         \\* \midrule
\endfirsthead
\multicolumn{6}{c}%
{{\bfseries Supplementary table 3 continued from previous page}} \\
\endhead
\bottomrule
\endfoot
\endlastfoot
\multicolumn{2}{l}{\textbf{Mar-20}} & \textbf{}           &                     &                     &                        \\
\textbf{}  & HFR                    & 1.4\% (0.4 - 5.7\%) & 4.0\% (1.7 - 9.4\%)   & 5.7\% (4.5 - 7.3\%) & 22.3\% (20.9 - 23.9\%) \\
\textbf{}  & Median LOS (death)     & 6.0 (2.9 - 7.8)       & 4.0 (1.4 - 5.9)       & 7.7 (7.1 - 8.3)     & 7.5 (7.3 - 7.9)        \\
\textbf{}  & Median LOS (discharge) & 1.0 (0.9 - 1.1)       & 2.4 (2.0 - 2.8)       & 3.0 (2.8 - 3.2)       & 5.6 (5.4 - 5.7)        \\
\multicolumn{2}{l}{\textbf{Apr-20}} &                     &                     &                     &                        \\
\textbf{}  & HFR                             & 0.5\% (0.1 - 3.9\%)  & 2.6\% (1.1 - 6.2\%) & 5.0\% (4.1 - 6.1\%)   & 18.2\% (17.3 - 19.2\%) \\
\textbf{}  & Median LOS (death)     & 5.0 (5.0 - 5.0)           & 4.7 (0.8 - 8.9)     & 5.3 (5.1 - 5.6)     & 6.7 (6.5 - 6.8)        \\
\textbf{}  & Median LOS (discharge) & 1.4 (1.1 - 1.7)     & 1.6 (1.3 - 1.9)     & 2.9 (2.8 - 3.1)     & 4.8 (4.7 - 4.8)        \\
\bottomrule
\end{longtable}

\begin{longtable}{@{}lllll@{}}
\caption*{{\itshape Supplementary table 3 continued}}
\label{tab:my-table}\\
\toprule
\multicolumn{2}{l}{\textbf{\begin{tabular}[c]{@{}l@{}}Month of hospital \\ admission\end{tabular}}} & \textbf{65-74}         & \textbf{75-84}         & \textbf{85+}           \\* \midrule
\endfirsthead
\multicolumn{5}{c}%
{{\bfseries Supplementary table 3 continued from previous page}} \\
\endhead
\bottomrule
\endfoot
\endlastfoot
\multicolumn{2}{l}{\textbf{Mar-20}} & \textbf{}              &                        &                        \\
\textbf{}  & HFR                    & 43.2\% (41.1 - 45.4\%) & 57.3\% (55.5 - 59.2\%) & 66.5\% (64.4 - 68.7\%) \\
\textbf{}  & Median LOS (death)     & 6.2 (6.0 - 6.4)          & 4.9 (4.8 - 5.0)          & 5.0 (4.8 - 5.1)          \\
\textbf{}  & Median LOS (discharge) & 6.8 (6.6 - 7.0)          & 7.8 (7.6 - 8.0)          & 9.5 (9.3 - 9.8)        \\
\multicolumn{2}{l}{\textbf{Apr-20}} &                        &                        &                        \\
\textbf{}  & HFR                              & 38.2\% (36.8 - 39.7\%) & 50.4\% (49.2 - 51.7\%) & 60.0\% (58.7 - 61.4\%)   \\
\textbf{}  & Median LOS (death)     & 5.6 (5.5 - 5.8)        & 4.8 (4.7 - 4.9)        & 4.9 (4.8 - 5.0)          \\
\textbf{}  & Median LOS (discharge) & 6.6 (6.5 - 6.7)        & 7.8 (7.7 - 7.9)        & 8.5 (8.4 - 8.7)        \\
\bottomrule
\end{longtable}

\endgroup

\cleardoublepage

\begingroup\small
\begin{longtable}{@{}llll@{}}
\caption*{Supplementary table 4: Hospitalised fatality risk and median length of stay (days) in hospital prior to death or discharge by month of hospital admission and sex (unadjusted for other covariates). Figures in brackets represent 95\% confidence intervals.}
\label{tab:my-table}\\
\toprule
\multicolumn{2}{l}{\textbf{Month of hospital admission}} & \textbf{Male} & \textbf{Female} \\* \midrule
\endfirsthead
\multicolumn{4}{c}%
{{\bfseries Supplementary table 4 continued from previous page}} \\
\endhead
\bottomrule
\endfoot
\endlastfoot
\multicolumn{2}{l}{\textbf{Mar-20}} & \textbf{}              & \textbf{}              \\
\textbf{}  & HFR                    & 43.7\% (42.5 - 44.9\%) & 35.2\% (33.8 - 36.6\%) \\
\textbf{}  & Median LOS (death)     & 5.8 (5.7 - 5.9)        & 5.2 (5.0 - 5.3)          \\
\textbf{}  & Median LOS (discharge) & 6.1 (6.0 - 6.2)          & 5.7 (5.5 - 5.8)        \\
\multicolumn{2}{l}{\textbf{Apr-20}} &                        &                        \\
           & HFR                    & 39.7\% (38.8 - 40.5\%) & 34.4\% (33.4 - 35.3\%) \\
\textbf{}  & Median LOS (death)     & 5.4 (5.3 - 5.4)        & 5.0 (4.9 - 5.1)          \\
\textbf{}  & Median LOS (discharge) & 5.8 (5.8 - 5.9)        & 5.8 (5.8 - 5.9)        \\
\bottomrule
\end{longtable}
\endgroup

\cleardoublepage

\begingroup\small
\begin{longtable}{@{}lllll@{}}
\caption*{Supplementary table 5: Hospitalised fatality risk and median length of stay (days) in hospital prior to death or discharge by month of hospital admission and ethnicity (unadjusted for other covariates). Figures in brackets represent 95\% confidence intervals.}
\label{tab:my-table}\\
\toprule
\multicolumn{2}{l}{\textbf{Month of hospital admission}} & \textbf{Asian} & \textbf{Black} & \textbf{White} \\* \midrule
\endfirsthead
\multicolumn{5}{c}%
{{\bfseries Supplementary table 5 continued from previous page}} \\
\endhead
\bottomrule
\endfoot
\endlastfoot
\multicolumn{2}{l}{\textbf{Mar-20}} & \textbf{}              & \textbf{}              & \textbf{}              \\
\textbf{}  & HFR                    & 36.7\% (34.1 - 39.4\%) & 35.8\% (33.1 - 38.6\%) & 42.3\% (41.2 - 43.4\%) \\
\textbf{}  & Median LOS (death)     & 5.6 (5.3 - 5.9)        & 6.2 (5.9 - 6.5)        & 5.4 (5.3 - 5.5)        \\
\textbf{}  & Median LOS (discharge) & 4.2 (4.1 - 4.4)        & 5.9 (5.8 - 6.1)        & 6.3 (6.2 - 6.4)        \\
\multicolumn{2}{l}{\textbf{Apr-20}} &                        &                        &                        \\
           & HFR                    & 28.8\% (27.0 - 30.7\%)   & 29.3\% (27.1 - 31.8\%) & 40.1\% (39.3 - 40.8\%) \\
\textbf{}  & Median LOS (death)     & 4.8 (4.6 - 5.0)          & 4.8 (4.6 - 5.0)          & 5.2 (5.2 - 5.3)        \\
\textbf{}  & Median LOS (discharge) & 4.1 (3.9 - 4.2)        & 5.0 (4.8 - 5.1)          & 6.3 (6.3 - 6.4)        \\
\bottomrule
\end{longtable}
\endgroup

\cleardoublepage

\begingroup\footnotesize
\begin{longtable}{@{}lllll@{}}
\caption*{Supplementary table 6: Hospitalised fatality risk and median length of stay (days) in hospital prior to death or discharge by month of hospital admission and region of residence (unadjusted for other covariates). Figures in brackets represent 95\% confidence intervals.}
\label{tab:my-table}\\
\toprule
\multicolumn{2}{l}{\textbf{Month of hospital admission}} & \textbf{London}      & \textbf{East Midlands} & \textbf{East of England} \\* \midrule
\endfirsthead
\multicolumn{5}{c}%
{{\bfseries Supplementary table 6 continued from previous page}} \\
\endhead
\bottomrule
\endfoot
\endlastfoot
\multicolumn{2}{l}{\textbf{Mar-20}} & \textbf{}              & \textbf{}              & \textbf{}              \\
\textbf{}  & HFR                    & 40.8\% (39.2 - 42.4\%) & 33.9\% (30.9 - 37.1\%) & 42.5\% (39.5 - 45.6\%) \\
\textbf{}  & Median LOS (death)     & 5.8 (5.7 - 6.0)          & 6.2 (5.9 - 6.5)        & 5.8 (5.6 - 6.1)        \\
\textbf{}  & Median LOS (discharge) & 6.1 (5.9 - 6.2)        & 6.0 (5.8 - 6.3)          & 5.6 (5.4 - 5.8)        \\
\multicolumn{2}{l}{\textbf{Apr-20}} &                        &                        &                        \\
\textbf{}  & HFR                            & 35.3\% (34.0 - 36.7\%) & 34.9\% (32.7 - 37.1\%) & 41.9\% (40.1 - 43.8\%)   \\
\textbf{}  & Median LOS (death)     & 4.9 (4.8 - 5.0)          & 5.0 (4.9 - 5.2)          & 5.3 (5.1 - 5.4)        \\
\textbf{}  & Median LOS (discharge) & 5.4 (5.3 - 5.5)        & 5.5 (5.4 - 5.7)        & 5.9 (5.8 - 6.1)        \\
\bottomrule
\end{longtable}

\begin{longtable}{@{}lllll@{}}
\caption*{{\itshape Supplementary table 6 continued}}
\label{tab:my-table}\\
\toprule
\multicolumn{2}{l}{\textbf{Month of hospital admission}} & \textbf{North East}    & \textbf{North West}    & \textbf{South East}    \\* \midrule
\endfirsthead
\multicolumn{5}{c}%
{{\bfseries Supplementary table 6 continued from previous page}} \\
\endhead
\bottomrule
\endfoot
\endlastfoot
\multicolumn{2}{l}{\textbf{Mar-20}} & \textbf{}              & \textbf{}              & \textbf{}              \\
\textbf{}  & HFR                    & 38.2\% (32.3 - 45.2\%) & 41.4\% (38.9 - 44.0\%)   & 38.2\% (35.7 - 41.0\%)   \\
\textbf{}  & Median LOS (death)     & 5.3 (5.0 - 5.7)          & 5.3 (5.1 - 5.6)        & 5.3 (5.0 - 5.5)          \\
\textbf{}  & Median LOS (discharge) & 6.6 (6.2 - 7.1)        & 6.3 (6.1 - 6.5)        & 6.2 (6.0 - 6.4)          \\
\multicolumn{2}{l}{\textbf{Apr-20}} &                        &                        &                        \\
\textbf{}  & HFR                             & 36.5\% (34.1 - 39.1\%) & 38.8\% (37.3 - 40.3\%) & 34.7\% (33.0 - 36.5\%)   \\
\textbf{}  & Median LOS (death)     & 4.9 (4.7 - 5.1)        & 5.3 (5.2 - 5.5)        & 5.5 (5.3 - 5.6)        \\
\textbf{}  & Median LOS (discharge) & 5.9 (5.8 - 6.1)        & 6.1 (6.0 - 6.2)          & 5.9 (5.8 - 6.1)        \\
\bottomrule
\end{longtable}

\begin{longtable}{@{}lllll@{}}
\caption*{{\itshape Supplementary table 6 continued}}
\label{tab:my-table}\\
\toprule
\multicolumn{2}{l}{\textbf{Month of hospital admission}} & \textbf{South West}    & \textbf{West Midlands} & \textbf{Yorkshire and Humber} \\* \midrule
\endfirsthead
\multicolumn{5}{c}%
{{\bfseries Supplementary table 6 continued from previous page}} \\
\endhead
\bottomrule
\endfoot
\endlastfoot
\multicolumn{2}{l}{\textbf{Mar-20}} & \textbf{}              & \textbf{}              & \textbf{}              \\
\textbf{}  & HFR                    & 37.6\% (33.8 - 41.8\%) & 42.0\% (39.4 - 44.7\%)   & 45.6\% (41.8 - 49.6\%) \\
\textbf{}  & Median LOS (death)     & 5.5 (5.2 - 5.9)        & 5.1 (4.9 - 5.3)        & 5.5 (5.2 - 5.9)        \\
\textbf{}  & Median LOS (discharge) & 5.2 (4.9 - 5.6)        & 4.7 (4.5 - 4.9)        & 6.3 (6.0 - 6.5)          \\
\multicolumn{2}{l}{\textbf{Apr-20}} &                        &                        &                        \\
\textbf{}  & HFR                         & 30.4\% (28.1 - 33.0\%)   & 39.0\% (37.1 - 41.0\%)     & 41.5\% (39.6 - 43.6\%)        \\
\textbf{}  & Median LOS (death)     & 5.3 (5.1 - 5.5)        & 5.3 (5.1 - 5.5)        & 5.4 (5.2 - 5.6)        \\
\textbf{}  & Median LOS (discharge) & 5.9 (5.7 - 6.1)        & 6.3 (6.1 - 6.4)        & 5.9 (5.8 - 6.1)        \\
\bottomrule
\end{longtable}

\endgroup

\cleardoublepage

\begingroup\footnotesize
\setlength{\tabcolsep}{2pt} 
\begin{longtable}{@{}llllll@{}}
\caption*{Supplementary table 7: Hospitalised fatality risk and median length of stay (days) in hospital prior to death or discharge by month of hospital admission and Charleson Comorbidity Index (unadjusted for other covariates). Figures in brackets represent 95\% confidence intervals.}
\label{tab:my-table}\\
\toprule
\multicolumn{2}{l}{\textbf{\begin{tabular}[c]{@{}l@{}}Month of hospital \\ admission\end{tabular}}} & \textbf{0}          & \textbf{1-2}           & \textbf{3-4}           & \textbf{5+}            \\* \midrule
\endfirsthead
\multicolumn{6}{c}%
{{\bfseries Supplementary table 7 continued from previous page}} \\
\endhead
\bottomrule
\endfoot
\endlastfoot
\multicolumn{2}{l}{\textbf{Mar-20}} & \textbf{}              & \textbf{}              & \textbf{}              & \textbf{}              \\
\textbf{}  & HFR                    & 17.4\% (16.1 - 18.8\%) & 36.3\% (35.0 - 37.8\%)   & 53.4\% (51.1 - 55.8\%) & 53.3\% (50.0 - 56.8\%)   \\
\textbf{}  & Median LOS (death)     & 6.0 (5.8 - 6.3)          & 5.6 (5.5 - 5.8)        & 4.9 (4.8 - 5.1)        & 5.0 (4.8 - 5.3)          \\
\textbf{}  & Median LOS (discharge) & 4.4 (4.3 - 4.5)        & 6.4 (6.3 - 6.5)        & 8.2 (8.0 - 8.4)          & 8.3 (8.0 - 8.6)          \\
\multicolumn{2}{l}{\textbf{Apr-20}} &                        &                        &                        &                        \\
           & HFR                    & 15.2\% (14.4 - 16.2\%) & 34.4\% (33.5 - 35.4\%) & 48.1\% (46.6 - 49.6\%) & 52.4\% (50.2 - 54.7\%) \\
\textbf{}  & Median LOS (death)     & 5.6 (5.5 - 5.7)        & 5.1 (5.0 - 5.2)          & 4.9 (4.8 - 5.0)          & 4.6 (4.5 - 4.8)        \\
\textbf{}  & Median LOS (discharge) & 4.3 (4.2 - 4.3)        & 6.2 (6.1 - 6.2)        & 7.9 (7.8 - 8.0)          & 8.2 (8.0 - 8.4)          \\
\bottomrule
\end{longtable}
\endgroup

\cleardoublepage

\begingroup\footnotesize
\begin{longtable}{@{}lllll@{}}
\caption*{Supplementary table 8: Hospitalised fatality risk and median length of stay (days) in hospital prior to death or discharge by month of hospital admission and measure of hospital load (unadjusted for other covariates). Estimates are replaced with a dash (-) where insufficient information was available. Figures in brackets represent 95\% confidence intervals.}
\label{tab:my-table}\\
\toprule
\multicolumn{2}{l}{\textbf{Month of hospital admission}} & \textbf{(0,20{]}}    & \textbf{(20,40{]}}     & \textbf{(40,60{]}}    \\* \midrule
\endfirsthead
\multicolumn{5}{c}%
{{\bfseries Supplementary table 8 continued from previous page}} \\
\endhead
\bottomrule
\endfoot
\endlastfoot
\multicolumn{2}{l}{\textbf{Mar-20}} & \textbf{}              & \textbf{}              & \textbf{}              \\
\textbf{}  & HFR                    & 38.8\% (36.6 - 41.0\%)   & 40.1\% (38.5 - 41.7\%) & 40.0\% (38.3 - 41.9\%)   \\
\textbf{}  & Median LOS (death)     & 6.1 (5.9 - 6.2)        & 5.7 (5.6 - 5.9)        & 5.2 (5.1 - 5.4)        \\
\textbf{}  & Median LOS (discharge) & 5.6 (5.5 - 5.8)        & 6.0 (5.9 - 6.1)          & 6.0 (5.9 - 6.2)          \\
\multicolumn{2}{l}{\textbf{Apr-20}} &                        &                        &                        \\
\textbf{}  & HFR                      & 34.3\% (32.6 - 36.0\%) & 37.3\% (36.2 - 38.4\%) & 37.8\% (36.6 - 39.0\%)  \\
\textbf{}  & Median LOS (death)     & 5.6 (5.5 - 5.8)        & 5.3 (5.2 - 5.4)        & 5.3 (5.2 - 5.4)        \\
\textbf{}  & Median LOS (discharge) & 6.0 (5.8 - 6.1)          & 5.9 (5.8 - 5.9)        & 5.7 (5.6 - 5.8)        \\
\bottomrule
\end{longtable}

\begin{longtable}{@{}lllll@{}}
\caption*{{\itshape Supplementary table 8 continued}}
\label{tab:my-table}\\
\toprule
\multicolumn{2}{l}{\textbf{Month of hospital admission}} & \textbf{(60,80{]}}   & \textbf{(80,90{]}}     & \textbf{(90,100{]}}    \\* \midrule
\endfirsthead
\multicolumn{5}{c}%
{{\bfseries Supplementary table 8 continued from previous page}} \\
\endhead
\bottomrule
\endfoot
\endlastfoot
\multicolumn{2}{l}{\textbf{Mar-20}} & \textbf{}              & \textbf{}              & \textbf{}              \\
\textbf{}  & HFR                    & 41.9\% (39.7 - 44.2\%) & 41.9\% (37.9 - 46.3\%) & 44.9\% (39.6 - 50.9\%) \\
\textbf{}  & Median LOS (death)     & 5.5 (5.3 - 5.6)        & 5.5 (5.2 - 5.9)        & 4.2 (3.8 - 4.7)        \\
\textbf{}  & Median LOS (discharge) & 5.8 (5.6 - 6.0)          & 6.5 (6.2 - 6.9)        & 5.0 (4.6 - 5.4)          \\
\multicolumn{2}{l}{\textbf{Apr-20}} &                        &                        &                        \\
\textbf{}  & HFR                             & 38.4\% (37.0 - 39.8\%) & 38.4\% (35.7 - 41.4\%) & 38.6\% (35.7 - 41.7\%) \\
\textbf{}  & Median LOS (death)     & 4.9 (4.8 - 5.0)          & 4.7 (4.4 - 4.9)        & 5.2 (4.9 - 5.4)        \\
\textbf{}  & Median LOS (discharge) & 5.7 (5.6 - 5.8)        & 5.7 (5.5 - 5.9)        & 6.9 (6.7 - 7.2)        \\
\bottomrule
\end{longtable}

\endgroup

\cleardoublepage

\begingroup\normalsize
\begin{longtable}{@{}ll@{}}
\caption*{Supplementary table 9: Hospitalised fatality sub-distribution hazard ratio by month of hospital admission, stratified by age group, region of residence and vaccination status, with regression adjustment (main effects) on sex, ethnicity, IMD quintile, CCI and hospital load. Figures in brackets represent 95\% confidence intervals.}
\label{tab:my-table}\\
\toprule
\textbf{Month of hospital admission} & \textbf{Hazard ratio} \\* \midrule
\endfirsthead
\multicolumn{2}{c}%
{{\bfseries Supplementary table 9 continued from previous page}} \\
\endhead
\bottomrule
\endfoot
\endlastfoot
Mar-20              & 2.30 (2.08, 2.55)                   \\
Apr-20              & 1.93 (1.75, 2.13)                  \\
May-20              & 1.33 (1.20, 1.48)                   \\
Jun-20              & 1 (reference category)             \\
Jul-20              & 0.74 (0.61, 0.91)                  \\
Aug-20              & 0.87 (0.69, 1.08)                  \\
Sep-20              & 1.07 (0.94, 1.21)                  \\
Oct-20              & 1.11 (1.00, 1.23)                     \\
Nov-20              & 1.14 (1.03, 1.26)                  \\
Dec-20              & 1.30 (1.17, 1.44)                   \\
Jan-21              & 1.32 (1.19, 1.46)                  \\
Feb-21              & 1.19 (1.07, 1.32)                  \\
Mar-21              & 1.00 (0.88, 1.14)                     \\
Apr-21              & 0.90 (0.75, 1.10)                    \\
May-21              & 1.20 (0.97, 1.50)                    \\
Jun-21              & 1.43 (1.23, 1.68)                  \\
Jul-21              & 1.31 (1.16, 1.47)                  \\
Aug-21              & 1.31 (1.17, 1.48)                  \\
Sep-21              & 1.15 (1.02, 1.30)                   \\* \bottomrule
\end{longtable}
\endgroup

\cleardoublepage

\begingroup\normalsize
\begin{longtable}{@{}lll@{}}
\caption*{Supplementary table 10: Hospitalised fatality sub-distribution hazard ratios by person characteristic, stratified by age group, PHE centre of residence and month of hospitalisation, with regression adjustment (main effects) on vaccination status, sex, ethnicity, IMD quintile, CCI and hospital load. January to September 2021. Figures in brackets represent 95\% confidence intervals.}
\label{tab:my-table}\\
\toprule
\multicolumn{2}{l}{\textbf{Characteristic}}                  & \textbf{Hazard ratio}  \\* \midrule
\endfirsthead
\multicolumn{3}{c}%
{{\bfseries Supplementary table 10 continued from previous page}} \\
\endhead
\bottomrule
\endfoot
\endlastfoot
\multicolumn{2}{l}{\textbf{Vaccination   status}}            & \textbf{}              \\
               & Unvaccinated                                & 1 (reference category) \\
          & \textless{}21 days after first vaccination dose          & 0.93 (0.89 - 0.98)         \\
               & $\geq$21 days after first vaccination dose       & 0.71 (0.67 - 0.77)     \\
               & $\geq$14 days after second vaccination dose      & 0.56 (0.52 - 0.61)     \\
\multicolumn{2}{l}{\textbf{Sex}}                             &                        \\
\textbf{}      & Male                                        & 1.28 (1.24 - 1.32)     \\
\textbf{}      & Female                                      & 1 (reference category) \\
\bottomrule
\end{longtable}
\endgroup

\cleardoublepage

\section*{Supplementary references}

\begin{enumerate}
    \item Andersen PK, Geskus RB, de Witte T, Putter H. Competing risks in epidemiology: possibilities and pitfalls. Int J Epidemiol. 2012;41: 861–870.
    \item Aalen OO, Johansen S. An Empirical Transition Matrix for Non-Homogeneous Markov Chains Based on Censored Observations. Scand Stat Theory Appl. 1978;5: 141–150.
    \item Fine JP, Gray RJ. A proportional hazards model for the subdistribution of a competing risk. J Am Stat Assoc. 1999.
    \item Bradburn MJ, Clark TG, Love SB, Altman DG. Survival analysis part II: multivariate data analysis--an introduction to concepts and methods. Br J Cancer. 2003;89: 431–436.
    \item Seaman SR, Nyberg T, Overton CE, Pascall D, Presanis AM, De Angelis D. Adjusting for time of infection or positive test when estimating the risk of a post-infection outcome in an epidemic. 2021. doi:10.1101/2021.08.13.21262014
\end{enumerate}

\end{document}